\documentclass[12pt, oneside, openright, a4paper]{book}

\usepackage[english]{babel}
\usepackage[dvips]{graphicx}
\DeclareGraphicsExtensions{.pdf,.png,.jpg,.ps}
\usepackage[utf8x]{inputenc}
\usepackage{amsfonts}
\usepackage{amssymb}
\usepackage{amsmath}
\usepackage{appendix}
\usepackage{color}
\usepackage{epsfig}
\usepackage{mathrsfs}
\usepackage[numbers, sort&compress]{natbib} 
\usepackage[font=footnotesize,hang]{caption} 
\usepackage{indentfirst} 
\usepackage{mathrsfs}
\usepackage{euscript}
\usepackage[bottom]{footmisc}
\usepackage[Lenny]{fncychap} 
\usepackage{minitoc} 
\mtcselectlanguage{english}
\usepackage[english]{varioref} 
\date{}

\usepackage{verbatim}
\usepackage{numprint} 
\usepackage{eso-pic}
\usepackage{subfigure} 
\usepackage{multirow} 
\usepackage{epigraph} 

\advance\textwidth by -0.7truecm
\advance\oddsidemargin by 0.4truecm
\advance\baselineskip by \baselineskip

\usepackage{fancyhdr}
\newcommand{\logo}[1]{\AddToShipoutPicture*{\AtPageCenter{\makebox(0,0){\includegraphics[width=0.6\paperwidth]{#1}}}}}

\begin{document}

\graphicspath{{figure/}} 

\selectlanguage{english}

\frontmatter


\logo{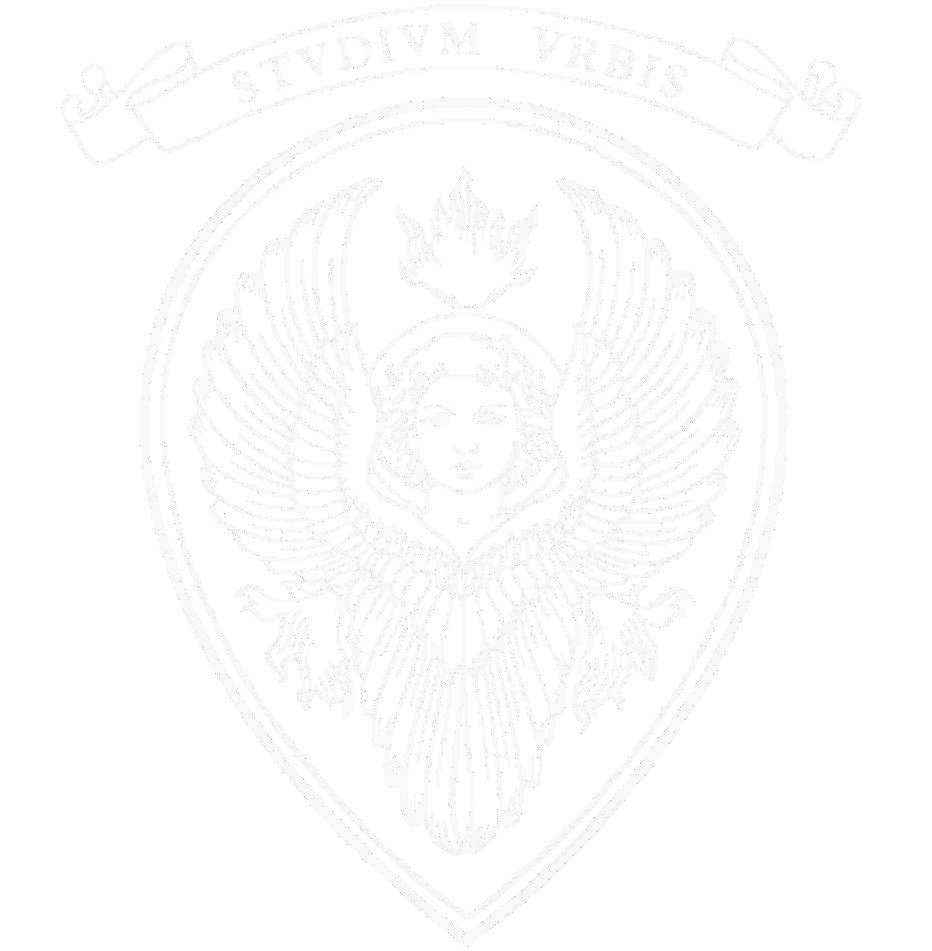}

\begin{titlepage}
\begin{center}

\uppercase{Universit\`a degli Studi di Roma, La Sapienza}\\
{\sc facolt\`a di scienze matematiche, fisiche e naturali}\\
{\sc corso di laurea specialistica in fisica}\\
\hbox to \textwidth{\hrulefill}

\vspace{1.5truecm}

{\sc Marco BAITY JESI}

\vspace{2truecm}

\sc{Energy Landscape in 3-Dimensional \\Heisenberg Spin Glasses}\\
\vspace{1truecm}
\sc{Scenario Energetico nei Vetri di Spin \\di Heisenberg 3-Dimensionali}\\

\vspace{2.5truecm}

\centerline{\hbox to 4.0truecm{\hrulefill}}

\medskip

{\sc tesi di laurea specialistica}\\

\centerline{\hbox to 5.5truecm{\hrulefill}}

\vspace{1truecm}

\begin{minipage}{\textwidth}
\begin{flushright}
\begin{minipage}{0.3\textwidth}
\begin{tabbing}
Chiar.mo \= Prof. Pinco Pallino \kill
Relatori:\>\\ 
         \> Prof. Federico RICCI TERSENGHI \\
         \> (\emph{Roma, La Sapienza})\\
         \> \\
         \> \\
         \> \\
         \> Prof. Victor MARTIN MAYOR\\
         \> (\emph{Madrid, Universidad Complutense})\\
\end{tabbing}
\end{minipage}
\end{flushright}
\end{minipage}

\vspace{1truecm}

\hbox to \textwidth{\hrulefill}
{\sc anno accademico 2009/2010}

\end{center}
\end{titlepage}


\setlength{\baselineskip}{1.35\baselineskip} 

\newpage\null\thispagestyle{empty}\newpage 
\thispagestyle{empty}
\begin{flushright}
\null\vspace{\stretch{1}}
{\emph{A mi mam\'a, \\
que me educ\'o como mejor no era posible.}}
\vspace{\stretch{2}}\null
\end{flushright}

\newpage\thispagestyle{empty}\null\newpage 

\dominitoc


\addcontentsline{toc}{chapter}{Abstract}
\chapter*{Abstract}
\markright{Abstract}{}

\paragraph{Italian Version}Recenti lavori suggeriscono che i vetri di spin di Heisenberg possano appartenere alla stessa classe di universalità di vetri 
strutturali. Trovare un equivalente su reticolo per questo tipo di modelli consentirebbe probabilmente di semplificarne lo studio sia dal punto di vista
numerico che analitico, consentendo eventualmente di rispondere alle molte domande riguardanti la transizione vetrosa che sono irrisolte da decenni.

I liquidi sottoraffreddati hanno comportamenti peculiari che, nel caso in cui l'analogia tra i due tipi di sistema reggesse, dovremmo riscontrare
nella fase paramagnetica dei vetri di spin di Heisenberg.

E' con questo obiettivo che nel presente lavoro si affronta in maniera sistematica la caratterizzazione della fase paramagnetica. Si cerca di 
comprendere il ruolo del panorama energetico, con uno studio approfondito
delle strutture inerenti (in analogia con i liquidi sottoraffreddati definiamo struttura inerente di una configurazione il minimo locale dell'energia 
più prossimo ad essa), che richiede lo sviluppo di un nuovo algoritmo di ricerca. Indaghiamo anche se si trovano tracce di una transizione dinamica
nella fase paramagnetica. Di fatto sia l'esistenza di uno scenario energetico complesso, sia quella di una transizione dinamica, sono tratti
caratteristici della fisica dei liquidi sottoraffreddati.

\newpage

\paragraph{English Version}Recent work suggests that Heisenberg spin glasses may belong to the same universality class than structural glasses.
Indeed, finding a lattice equivalent for supercooled liquids would probably allow easier numerical and analytical studies, that may help to answer 
long standing questions on the glass transition.
Supercooled liquids have many peculiar behaviors that should be found in the paramagnetic phase of Heisenberg spin glasses if the analogy 
between the two systems holds. 

It is with this motivation that we undertake a study of the paramagnetic phase of Heisenberg spin glasses. We shall emphasize the
role of the energy landscape, with a detailed study of the properties of the inherent structures (by analogy with supercooled liquids, we
name inherent structure the local minimum of the energy function which is closest to the current spin configuration). Finding inherent
structures will require the development of a new search algorithm. We shall investigate as well the existence of a dynamic "phase
transition" in the paramagnetic phase. As a matter of fact, both the existence of a complex energy landscape, as well as the existence of a
dynamic transition, are distinguished features of the Physics of supercooled liquids.

\adjustmtc[1]


\tableofcontents

\newpage
\thispagestyle{empty}
\null
\newpage

\mainmatter


\addcontentsline{toc}{chapter}{Introduction}
\chapter*{Introduction}
\markright{Introduction}{}

Spin and structural glasses are two completely different types of systems. Nevertheless, deep analogies can be found if 
one considers some dynamic properties, metastability, and the role of competing interactions in both types of systems.
The main aim of this thesis is to try to give a good starting point for a solid correspondence between those two
types of models. A similar job has been done with non-disordered systems, by making a one-to-one mapping between
the Ising model and lattice gases. Although they represent systems that apparently have nothing in common, those 
two models have the exact same behavior if we recognize the correct relation between the quantities that appear in 
each model.

Establishing a similar analogy between spin and structural glasses would probably be mutually beneficial for their study.
In particular for spin models it is easier to demonstrate theorems, that could extend also
to structural glasses and supercooled liquids. Besides, numerous approximation methods have been developed for spin glasses,
and the numerical study of spin glasses is far more manageable than the one of structural glasses,
since it is possible to apply very performing Monte Carlo algorithms on lattices, so that finding an eventual symmetry
relating the former with the latter could make us extend the results of one model to the other, and vice versa.

Although it is several decades that supercooled liquids and structural glasses are studied \cite{cavagna}, their comprehension is far from being
complete. In particular it is still not known if there exists a real thermodynamic glassy phase at low temperature. A salient feature 
is that lowering the temperature there is a very steep growth of the characteristic time scales $\tau$ of the system. 
The experimental approach is to define $T_{g}$ as the temperature at which $\tau\sim10^2-10^3s$, but this temperature has only a practical
meaning. 
From the theoretical side, the Mode Coupling Theory (MCT) \cite{MCT} offers a framework to
describe quantitatively this dynamic slowdown. In fact, Mode Coupling
Theory predicts a dynamic transition where the self-correlation times
diverge with power law. Nowadays, the Mode Coupling Transition is
known to be a crossover rather than a real transition. Nevertheless,
the functional forms predicted by MCT are routinely used to fit
experimental results when the autocorrelation times are
in the range $10^{-13}$ s $< \tau < 10^{-3}$ s.

Recently, a topological interpretation has been given for the
dynamic transition predicted by the MCT in which the role of the energy landscape, i.e. the topological aspect of the potential energy,
assumes a determinant role in the dynamics of these systems. At high temperatures the typical configurations in which the system
finds itself are closer to saddle points of the potential energy than minima, so there is always a negative direction in its hessian that permits
to the system a fast movement in the configuration space. Decreasing the temperature the typical number of negative directions
of these saddles starts falling too, so that the relaxation times increases as the system has always more difficulties in
finding these more convenient directions. There is a temperature in which the density of negative directions of the hessian
goes to zero. This change in the topology of the energy landscape is called topological transition, and it causes the dynamic transition, 
i.e. a high slowing down of the dynamics (but with no divergence and no real phase transition). 
Mode Coupling Theory predicts that the relaxation times diverge with power law at this temperature. This is not what happens
because when the saddles disappear we enter another regime of dynamics in which the system has to `jump' from one minimum to the other, 
in which the typical times are much longer since they scale with the exponential of the energy barrier. We can so appreciate the divergence
predicted by Mode Coupling Theory only until its characteristic time scales become comparable with those of activated dynamics.
Another interesting feature that follows from the energy landscape scenario is the concept of inherent structure. It is possible
to associate, to any state, the configuration corresponding to the local minimum where the system would find itself
if the energy were to be lowered abruptly in some way. We call that configuration inherent structure. What is seen is that at high temperatures
the energies of the inherent structures almost does not depend on $T$. On the contrary, at lower temperatures these energy suddenly start to 
decrease fairly quickly. The interpretation of this phenomenon is that there is an exponentially high number of minima at a certain level, and 
few with lower energy, so as long as we are at high temperature the probability of finding an exponentially numerous minimum is almost one,
and to feel the presence of the lower energy inherent structures we need to be at lower temperatures.

On the other hand, we have a stronger hold on the physics of
spin-glasses (the second partner in our intended analogy), specially
for Ising spin glasses.  Spin glasses \cite{spin-glass-theory,dotsenko} are disordered magnetic alloys,
such as Mn$_x$Cu$_{1-x}$, where magnetic moments (localized on the Mn
atoms, typically) interact. If these magnetic moments may point
basically anywhere, their interaction depending only on their relative
orientation, one speaks of Heisenberg spin glasses. Yet, if the
lattice anisotropies enforce the magnetic moments to point mostly to
the North or South poles, we have Ising spin glasses.

It is since the beginning of the $90$'s that there
is clear evidence of a thermodynamic spin glass phase at low temperatures for systems with Ising spins and finite-range interactions.
Also, there is no dynamic transition, in the sense that there isn't any temperature in the paramagnetic phase in which the relaxation
times grow considerably. Moreover, Parisi's solution of the mean field model yields an infinite-step replica symmetry breaking, that implies 
a continuous sequence of phase transitions once the threshold of $T_{SG}$ is passed, with a continuous fragmentation of the valleys of the 
free energy, and a macroscopically large number of different configurations that characterize the glassy phase. The slow down of the dynamics in 
the glassy phase would be due to this complexity of the phase space. It appears clear that Ising spins don't offer the possibility of a 
correspondence with structural glasses. 

In spite of the fact that they are a much more reasonable representation of real spin glasses, extremely little is known about Heisenberg spins 
glasses.  Although it is assumed that a phase transition takes place at finite temperature, its nature is not clear yet, since there is a degree 
of freedom that was absent in Ising spins, called chirality, that represents the handedness of the non-collinear orientation, and that could yield 
another phase transition. It is reason of intense debate whether spin and chirality transitions are two distinct phenomena, or if they take place 
at the same temperature. Not much more has been investigated, so it is worth to make an investigation to see if Heisenberg spin glasses have features
in common with structural glasses.

There are some reasons to conceive an analogy between structural
glasses and some types of spin-glasses. In 1987, Kirkpatrick and
Thriumalai \cite{pspinMCT} pointed out the formal identity between the Mode Coupling
equations and the exact equations for the dynamics of a particular
type of spin-glasses, with p-body interactions (the so called p-spin
models). Unfortunately, this analogy has been established so far only
for models with long range interactions, where the mean field
approximation is actually exact.
Inspired by this intuition, in the year $2000$ G.Parisi and M.M\'{e}zard \cite{parisimezard} made a replica study of structural glasses 
for finite-size systems that suggested that the structural
glass transition belongs to the same Universality Class of the systems
with one step of replica symmetry breaking.
That means that the order parameter's distribution is not continuous as in the Ising spin glasses, but it is takes only two values \footnote{In 
absence of spacial symmetries we have two peaks: one at zero, and one at a finite value $q_{EA}$. If, the system is symmetric under reflection 
there are three peaks because we have the reflection of $q_{EA}$, $-q_{EA}$}, because there is only one step of fragmentation of the free energy 
landscape. Unfortunately, a simple model, with short range interactions,
undergoing a 1-RSB glass transition, is yet to be found.

There are a couple of clues that make us think that finite-range interaction Heisenberg spin glasses are a good candidate for the correspondence with
spin glasses that we have exposed. The first is that H.Kawamura in $2005$ \cite{pqchir} found a double-peak distribution of the order parameter typical of the one-step 
replica symmetry breaking, so the universality class may be the same. The second is that up to our knowledge the only dynamic study of the paramagnetic phase done up to now
\cite{picco} estimates the phase transition at a temperature $50\%$ higher than its actual position, and in an optimistic scenario this overestimation
could be explained if what they found was actually the dynamic transition predicted by Mode Coupling Theory.

The above mentioned clues clearly do not suffice to establish
our seeked analogy. Hence, if one wants to have stronger evidence of the similarities between the two models it is necessary to make
a systematic research of the aspects under which we would like to remark the analogy. We conducted our research under two main aspects:
\begin{enumerate}
 \item Since the Mode Coupling Transition is best understood through the topological properties of the energy landscape,
 it is absolutely natural to start our investigation
 concentrating ourselves on the energy landscape of the Heisenberg spin glasses, so we disposed ourselves to a systematic research of inherent 
 structures. This type of investigation has never been done, because Ising spins are discrete variables, and Heisenberg spin glasses are still a 
 largely unexplored topic. The only energy-lowering routine used in bibliography is the quench, which is known to be very badly performing. Hence 
 we had to find a better performing algorithm to find inherent structures, and we made a deep study of their spatial correlation functions. We 
 succeeded in finding a fast algorithm that permitted us to make our research with very large samples, compatible with the thermodynamic limit, and 
 we noticed that the inherent structures are indeed relevant for the physics. First, we found the anharmonic
 term was rather small, so we can explain the energy of a thermalized state as harmonic fluctuations around a minimum. Second, lowering the temperature
 the correlation lengths of thermal states and the relative inherent structures converge.

 \item Mode Coupling Theory predicts a divergence of the relaxation times in the paramagnetic phase, so we investigated the equilibrium dynamics of 
 the Heisenberg spin glass model to see if we would encounter the same behavior. In conformity with the former point, we wanted to be in the thermodynamic
 limit, so we had to use non-physical algorithms to obtain thermalized configurations, and then we switched to physical dynamics. We wanted to verify 
 if the relaxation time admitted a power law scaling for $3$-$4$ decades that predicts a transition significantly over the critical temperature, 
 as it happens in structural glasses. Also this analysis was successful: we found a scaling with exponents of the same order, in the same range
 of times, and the estimate of the possible dynamic transition was consistently over the spin glass transition, and two standard deviations under the
 previous estimate given in \cite{picco}.
\end{enumerate}

After these results the analogy between the two systems looks definitely more solid. Stronger data for this hypothesis could be given by an analysis
of the variation with temperature of the density of negative directions in the hessian, which is a work actually in progress, and of the configurational
entropy.

\adjustmtc[1]



\newpage\thispagestyle{empty}\null\newpage 

\chapter{A brief summary on spin glasses}
\label{chap:one}
\minitoc
\mtcskip


\newpage

\section{Defining a spin glass model}
\label{sec:average}
In the last decades particular attention has been given to a set of models which were
meant to describe random magnets with co-presence and competition of ferromagnetic and
anti-ferromagnetic interactions.

\begin{figure}[!hb]
 \begin{center}
  \includegraphics{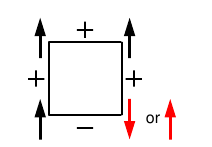}
  \caption{A toy model which shows frustration. If the interaction on the bond is a ``+'', the spins 
  want to be parallel and if it is a ``-'' they want to be anti-parallel. 
  The model is frustrated because it is impossible to satisfy all the conditions simultaneously. 
  In a spin lattice any loop of spins, in which the product of the interactions is negative, is frustrated.}
  \label{fig:frustration} 
 \end{center}
\end{figure}

From the pure modelisation point of view a spin glass has two key ingredients: disorder
and frustration. A simple example is given in figure \ref{fig:frustration}, where we have 
$4$ Ising spins (they can point only up or down) interacting in a loop. The interactions
between the spins are randomly ferromagnetic or anti-ferromagnetic, so it can happen, as in this case,
that it is impossible to satisfy all constraints. It is clear that in a very large system
finding the configuration of spins that satisfies the most number of constraints is a very 
hard task.

The first spin glass model was presented in 1975 by S.F. Edwards and P.W. Anderson \cite{EA}
to modelize a dilute solution of Mn in Cu, in which the interactions between particles can be
ferromagnetic or anti-ferromagnetic depending on their mutual distance.

In the Edwards-Anderson (EA) model the spins are placed in fixed positions in a $D$-dimensional
regular lattice. This is because the typical time scales of the movement of the particles is
overwhelmingly greater, and non-trivial interesting effects can be seen on a regular lattice.

The model's Hamiltonian is
\begin{equation}
 H=-\displaystyle\sum_{<\vec x\vec y>} J_{xy} \vec s_{\vec x} \cdot \vec s_{\vec y}
,
\end{equation}
where the brackets $<\cdot>$ indicate that the sum is performed only over the nearest neighbours.
The spins $\vec s_{\vec x}$ are $m$-dimensional vectors with $||\vec s_{\vec x}||=1$. The type of spins has a 
different name depending on $m$'s value:
\begin{itemize}
 \item $m=1$: Ising spins
 \item $m=2$: XY spins
 \item $m=3$: Heisenberg spins.
\end{itemize}
It is clear that the choice of $m$ depends strictly on the anisotropy of the material we want to study.
The couplings $J_{xy}$ are random variables with zero mean to avoid any bias towards ferromagnetism or antiferromagnetism,
and their standard deviation has to be finite. If we call $[\cdot]$ the average over the different
realizations over the disorder these conditions express as
\begin{equation}
 [J_{xy}]=0, ~~~~~~~~~~ [J_{xy}^2]^{1/2}=J(=1).
\end{equation}
The most used probability distribution functions for the $J_{xy}$'s are the bimodal and the Gaussian distribution.
We stress that in these systems disorder is not only due, like in most models, to thermal fluctuations, but it also appears
explicitly in the Hamiltonian, giving two levels of disorder. We decided to write the averages over the different
realizations of the couplings with square brackets $[\cdot]$, and the thermal averages will be written in angular brackets $<\cdot>$.

Experimentally we can find spin glasses in many different systems:
\begin{itemize}
 \item Metals:
Magnetic alloys in which we have atoms with strong mutual interactions diluted in a non-magnetic metal such as Mn in Cu have
two-site interactions
\begin{equation}
 J_{xy}\sim \frac{\cos(2k ||{\vec x}-{\vec y}||)}{||{\vec x}-{\vec y}||^3}
\end{equation}
which depend on the distance and evidently yield the possibility of having both ferromagnetic and anti-ferromagnetic interactions.
 \item Insulators:
  An example is Fe$_{0.5}$Mn$_{0.5}$TiO$_3$, which comprises hexagonal layers. The spins align perpendicular to the layers. Within a layer
 the spins in pure FeTiO$_3$ are ferromagnetically coupled while spins in MnTiO$_3$ are antiferromagnetically coupled. Hence the mixture 
 gives an Ising spin glass with short range interactions.
 \item Non-conventional spin glasses:
Spin glass theory has been used to study a numerous amount of problems outside the condensed matter domain, such as:
\begin{itemize}
 \item Optimization problems in computer science
 \item Protein folding
 \item Polymer glasses, foams...
\end{itemize}
\end{itemize}

\section{Order Parameters}
A phase transition presents often a spontaneous symmetry breaking in the system. The order parameter is the observable to whom we impute the breaking
of the symmetry. It usually transforms, under the symmetry changes, in a homogeneous non-trivial way, so that, when it is zero,
symmetry is preserved and, when it is non-zero, it is broken.

It is now assumed, after consistent experimental, theoretical and numerical work, that EA spin glasses have two thermodynamic phases:
the paramagnetic phase and the spin glass phase. This issue has been quite controversial for years, because the static aspect of a configuration
in the spin glass phase was not easily distinguishable from a paramagnetic one. As a matter of fact the order parameter of a spin glass is
not a trivial object, and it deserves particular attention.

\subsection{Ising Spins}
To make the order parameter definition easier we rely on the comparison with the order parameters in the non-disordered counterparts of our models.
Thus, for Ising spins we will show how the order parameter is defined in the Ising model, and see how to extend this definition to the
EA Ising model.

\subsubsection{Symmetries and Order Parameter in the Ising Model}
The Ising model's Hamiltonian is
\begin{equation}
 H=-\displaystyle\sum_{<\vec x\vec y>} s_{\vec x} s_{\vec y}
,
\end{equation}
with $s_{\vec x}=\pm 1$.

It is immediate to see that its symmetry group is $Z_2$, since making the symmetry transformation
\begin{equation}
 Z_2:~~ \{s_{\vec x}\} \rightarrow \{-s_{\vec x}\}
\end{equation}
the Hamiltonian and the functional integration measure are invariant.
The order parameter is the magnetization
\begin{equation}
 m=\frac{1}{N}\displaystyle\sum_{\vec x} s_{\vec x}
\end{equation}
which is not invariant under our $Z_2$ symmetry
\begin{equation}
 Z_2:~~ m \rightarrow -m
\end{equation}
so in the disordered phase, for $T>T_c$, we have $m=0$, and for $T<T_c$, in the ordered phase, we have $m\neq0$.
Unfortunately the symmetry breaking, hence the phase transition, occurs only in the thermodynamic limit $L\rightarrow\infty$ because in finite
systems the average magnetization is zero also in the ordered phase, since the energy barriers that cause the ergodicity breaking are infinite
only in an infinitely large system.
Therefore, if we want to extract infinite-size properties from a finite-size system, as it is done in numerical simulations, we
should better study invariant quantities that are not negatively influenced by the absence of phase transition.

The correlation function
\begin{equation}
 G(\vec x, \vec y) = < s_{\vec x} s_{\vec y}>
\end{equation}
is clearly invariant under our symmetry transformation
\begin{equation}
 Z_2:~~s_{\vec x} s_{\vec y}\rightarrow (-s_{\vec x})(- s_{\vec y})=s_{\vec x} s_{\vec y}
\end{equation}

The correlation function is strictly binded to the order parameter, since
\begin{equation}
 <m^2>~=~ <(\frac{1}{N}\displaystyle\sum_{\vec x} s_{\vec x})(\frac{1}{N}\displaystyle\sum_{\vec y} s_{\vec y})>~=~\frac{1}{N^2}\displaystyle\sum_{\vec x}\sum_{\vec y} <s_{\vec x}s_{\vec y}>
.
\end{equation}
If the system has periodic boundary conditions the correlation function is also translationally invariant so $G(\vec x, \vec y)=G(\vec x-\vec y)$, hence
\begin{equation}
 <m^2>~=~ \frac{1}{N}\displaystyle\sum_{\vec r}<s_0 s_{\vec r}>~=~\frac{1}{N}\displaystyle\sum_{\vec r}G(\vec r)~=~\frac{1}{N} \hat{G}(\vec k =\vec 0)
\end{equation}
where the hat indicates the function's Fourier transform. We see then that the correlation function contains a lot of interesting information over the thermodynamic system.

\subsubsection{Symmetries and Order Parameter in the EA Ising Model}
\label{sec:IsingOP}
Introducing disorder in the Ising model we have
\begin{equation}
 H=-\displaystyle\sum_{<\vec x\vec y>} J_{\vec x \vec y} s_{\vec x} s_{\vec y}
\end{equation}
which, as we said in section \ref{sec:average}, puts us in the position to have to define two levels of disorder.
We have two types of symmetry now. One is the global $Z_2$ related to the positions of the spins belonging to a single sample
of which we just talked. The second is a local symmetry (gauge symmetry) related to the realizations of the disorder.
Speaking more clearly if we associate to each site $\vec x$ a variable $\eta_{\vec x}=\pm1$ it is clear that the transformation
\begin{eqnarray}
  J_{xy} && \rightarrow J_{xy}'= \eta_x J_{xy} \eta_y \\
  s_{x} && \rightarrow s_x'= \eta_x s_x \\
\end{eqnarray}
 keeps invariant the relevant magnitudes $H$ and $Z$. Notice that the probability distribution function of the $J$'s does not change
with the gauge transformation.

While the global invariance is verified only in the thermodynamic limit, the local symmetry is always true.
Let's see how it is possible to study a finite size system.


Since we want to study the properties of the system no matter the disorder realization, we can no longer use $s_xs_y$ as
observable for its investigation because it is not gauge-invariant.

To construct a gauge-invariant observable that preserves all the symmetries that we need, we resort to a simple trick,
redefining the Hamiltonian in a harmless way, in order for it to represent two replicas $(a)$ and $(b)$ of the same realization of the disorder:
\begin{equation}
 H^{(2)} = - \displaystyle\sum_{<x,y>}J_{xy} ( s^{(a)}_xs^{(a)}_y + s^{(b)}_xs^{(b)}_y )
\end{equation}
The local symmetry remains the same, while the global symmetry group changes then to $Z_2 \otimes Z_2$ because the invariance holds also if we flip all the spins of only one
of the two replicas.

We can now construct a gauge-invariant object
\begin{equation}
 q_x = s^{(a)}_xs^{(b)}_x
\end{equation}
that is not invariant under the global symmetry. If we define the Edwards-Anderson parameter as
\begin{equation}
 q = \frac{1}{N}\displaystyle\sum_x q_x
\end{equation}
we see that it is a good order parameter since under the global symmetry $q\rightarrow\pm q$.

As we did before we relate the square order parameter with the correlation function, which in presence of quenched disorder 
and translation invariance is
\begin{equation}
 G_J(x,y;J_{xy}) = [<q_xq_y>] = G_J(x-y;J_{xy})
,
\end{equation}
so
\begin{equation}
 [<q^2>]~=~(\frac{1}{N^2}\displaystyle\sum_x\sum_y [<q_xq_y>] )~=~\frac{1}{N}\sum_r G(r) ~=~ \frac{1}{N}\chi
.
\end{equation}

\subsection{Heisenberg Spins}
We have seen how to define the order parameter in the EA Ising model and how it is related to some other useful observable.
Is there a great difference when we pass from $m=1$ spin dimensions to $m=3$ ?

Indeed we know that the universality class will not be the same, so critical exponents and scaling functions will not be the same.
Nevertheless those are not the only important differences between the two models. It will be in the next chapter that we will
focus on the recent results that stress the diversity implied by $m=3$.

Here we will concentrate on the symmetries implied by $3$-dimensional spins, showing how it is possible to identify a new order
parameter that possibly changes the whole physics of the EA model.

\subsubsection{Symmetries and Order parameter in the Heisenberg Model}
In the Heisenberg model the Hamiltonian is slightly modified
\begin{equation}
 H=-\displaystyle\sum_{<x,y>} \vec s_x \cdot \vec s_y
\end{equation}
where now the spins $\vec s_x$ are $3$-dimensional vectors with the constraint $\vec s^2=1$. A Heisenberg system is invariant under
the rotation or inversion of all the spins, so its global symmetry group is 
\begin{equation}
O(3):~~\vec s_x \rightarrow T\vec s_x~~~~~,~~~~~TT^t=\mathbb{I}
\end{equation}
and the order parameter 
\begin{equation}
 \vec m = \frac{1}{N}\displaystyle\sum_x \vec s_x
\end{equation}
is vectorial.
As in the Ising case, for a finite system $<\vec m>=\vec 0$ at all temperatures, so there is no point in computing it, and we have to find
a non trivial quantity in finite volumes, as the correlation function
\begin{equation}
G(x,y)=<\vec s_x\cdot\vec s_y>
\end{equation}
that is related to the order parameter in the usual way
\begin{equation}
 <\vec m^2> = \frac{1}{N}\displaystyle\sum_r G(r)
\end{equation}

\subsubsection{Symmetries and Order parameter in the EA Heisenberg Model}
Knowing already the consequences of the presence of disorder we apply directly the trick seen in section \ref{sec:IsingOP} and we write the
two-replica Hamiltonian
\begin{equation}
 H=-\displaystyle\sum_{<x,y>} J_{xy} (\vec s^{(a)}_x \cdot \vec s^{(a)}_y + \vec s^{(b)}_x \cdot \vec s^{(b)}_y)
\end{equation}
that makes us possible to get the Edwards-Anderson parameter.
The symmetries are
\begin{itemize}
 \item Disorder:\hspace{2cm}$Z_2$ \hspace{0.5cm}(local)
 \item Thermal:\hspace{2cm}$O(3)\otimes O(3)$ \hspace{0.5cm}(global)
\end{itemize}
Our gauge invariant this time is a tensor that we call spin-glass site overlap
\begin{equation}
 \tau_{\alpha\beta}(\vec x) = \vec s^{(a)}_{\vec x,\alpha} \vec s^{(b)}_{\vec x,\beta}
,
\end{equation}
where $\alpha,\beta=1,2,3$ are the components of the spin vectors, thus the Edwards-Anderson parameter in the EA Heisenberg model is
\begin{equation}
 Q_{\alpha\beta} = \displaystyle\sum_x \tau_{\alpha\beta}(\vec x)
,
\end{equation}
that is related with the correlation function 
\begin{equation}
 G(x,y) = [<tr (\tau(\vec x)\tau^{\dagger}(\vec y))>]
\end{equation}
in the usual way
\begin{equation}
 [<tr(Q Q^{\dagger})>]~=~\frac{1}{N}\sum_{\vec r} G(\vec r) ~=~\frac{1}{N}\sum_{\vec r} \hat{G}(\vec k=0)~=~ \frac{1}{N}\chi
\end{equation}

\paragraph{Chirality}
The order parameter scenario in the EA Heisenberg model even more interesting than what we have just exposed. In fact one can admit the
possibility of breaking the global symmetry $Z_2$ keeping intact the $SO(3)$ symmetry \cite{kawamura:02}. 

We define chirality of a site along the direction $\mu=1,2,3$ as
\begin{equation}
 \kappa^\mu(\vec x) = \vec s_{\vec x - \hat e_\mu} \cdot (\vec s_{\vec x} \times \vec s_{\vec x + \hat e_\mu})
\end{equation}
where $\hat e_\mu$ is the unit vector in the direction $\mu$. If we make a mirror exchange of all the particles respect to
a plane the chirality of the sites remains the same except for the spins on the plane. A ferromagnetic system would not
feel the reflection because the system is homogeneous and non zero chirality is generated only by thermal fluctuations, and 
not by the energy landscape, so it would not be a good observable. On the other side spin glasses are not homogeneous, and
the in the ground states the spins are not aligned, so they are not invariant under this transformation.

In compliance with what we did above, we define our chiral gauge invariant
\begin{equation}
\epsilon^\mu (\vec x)= \kappa^{(a)}_\mu(\vec x)\kappa^{(b)}_\mu(\vec x)
, 
\end{equation}
from which descends the chiral-glass (CG) overlap in the direction $\mu$, which is our order parameter,
\begin{equation}
E^\mu = \frac{1}{N}\displaystyle\sum_{\vec x}\epsilon^\mu(\vec x)
,
\end{equation}
and we remark its connection with the simmetrized correlation function
\begin{equation}
 C^{CG}_{\mu\nu}(\vec x \vec y) = \frac{[<\epsilon^\mu (\vec x)\epsilon^\nu (\vec y)>+<\epsilon^\nu (\vec x)\epsilon^\mu (\vec y)>]}{2}
\end{equation}
through the usual relation
\begin{equation}
 [<E_\mu E_\nu>] ~=~ \frac{1}{N}\displaystyle\sum_{\vec r}C^{CG}_{\mu\nu}(\vec r) ~=~ \frac{1}{N}\chi_{CG}
\end{equation}

\paragraph{Overlaps}
A very handy feature of disordered systems is the overlap, a magnitude that tells us how similar two configurations are, 
being equal to zero if there is no resemblance, and far from zero in the opposite case.
There are numerous ways to define an overlap, but all of them are not invariant under the global symmetries of a Heisenberg 
spin glass. This could be a problem in numerical simulations, because we could think that two identical replicas, who differ 
only by a rotation, have very different configurations. Solutions to this would be very computationally expensive, so a
 symmetry-invariant definition of the overlap would be very useful.

For the chirality the solution is direct, since it is a scalar object and it is always invariant, so we can use 
the chiral glass overlap $E^\mu$ introduced just above.

For the spin-glass (SG) overlap, for example we could use the Edwards-Anderson parameter, but it is not $O(3)$ invariant.
If we look at the previous paragraphs we can remark that the square overlap
\begin{equation}
 Q^2 = tr (\tau(x)\tau^\dagger(x))
\end{equation}
not only is an invariant, but it also gives an idea of the similarity between two configurations. It is zero for completely
different replicas, and about $1/3$ when we compute it for a replica with itself and its spins are isotropically distributed
(in presence of anisotropy it tends to $1$).


\chapter{Update on developments in spin glasses}

\minitoc
\mtcskip

\newpage
\section{Ising vs Heisenberg}
The original version of the EA model was proposed for Heisenberg spins, as it is expected that they represent a good approximation for
most of the spin glass material, in particular those who posses weak magnetic anisotropy. In fact Heisenberg spins are much more present in 
nature than Ising spins, so it is logical to be more interested in them.Nevertheless Ising spins have been studied
far more because of their greater simplicity. It has been argued, in the initial studies of the EA Ising model, that their weak magnetic 
anisotropy present in real magnets causes a crossover from the isotropic behavior to the anisotropic, shifting the universality
class of the systems, so that Ising spins, corresponding to the strong anisotropy limit would describe also the behavior of
magnets with weak anisotropy \cite{bray}.

\subsection{Ising ordering in zero field}
It hasn't been clear whether the three dimensional EA Ising model presented a finite-temperature spin-glass transition until in 1985
Monte Carlo simulations from Ogielski, Bhatt and Young \cite{tsg1,tsg2} found solid evidence of a finite $T_{SG}$. Later investigations
on the critical exponents found values that were difficult to compare with an experimental realization of the Ising spin glass because
at present we know of only one type of Ising spin glass really existing in nature that we can study in laboratory, an insulating spin
glass magnet of Fe$_{0.5}$Mn$_{0.5}$TiO$_3$ that we have already mentioned in section \ref{sec:average}. We stress that when estimates of the 
critical exponents started to be found the argument mentioned above, that Ising spins may fairly represent a weakly anisotropic
Heisenberg spin glass, collapsed immediately, since they were very far from being compatible.

\begin{figure}[!ht]
 \begin{center}
  \includegraphics[width=\textwidth]{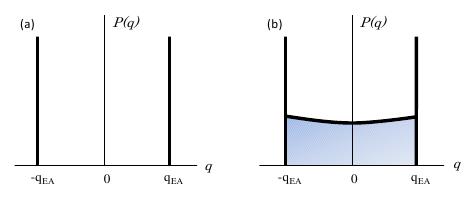}
  \caption{The overlap distribution $P(q)$ in the thermodynamic limit for a finite-range spin glass. In figure $(a)$ we show the droplet picture: the spin glass is a
  disguised ferromagnet and has only two ordered states, represented with two delta functions at $\pm q_{EA}$. In figure $(b)$ we see how with hierarchical
  RSB $P(q)$ possesses an additional plateau connecting the two delta-function peaks.}
  \label{fig:P_q}
 \end{center}
\end{figure}

A point that raises a lot of controversies is the nature of the SG ordered state. There are two dominant visions: the ``droplet picture'' that
claims that EA spin glasses are disguised ferromagnets without a spontaneous Replica Symmetry Breaking (RSB) and only two ordered states \cite{droplet}, 
and the ``hierarchical RSB picture'', inspired on Parisi's exact solution of the mean field model, which claims that the ordered state is accompanied
with a hierarchical RSB \cite{RSB}. In figure \ref{fig:P_q} we show the distribution of the order parameter in one case and in the other.

\subsection{Heisenberg ordering in zero field and chirality scenario}

Early numerical simulations individuated in common agreement that the EA Heisenberg model had only a zero temperature phase transition, in evident contrast with
experiments, which clearly showed a finite $T_{SG}$. The scientific community used to explain this inconsistency invoking the weak magnetic anisotropy present 
in real spin glasses, that wasn't reproduced in numerical simulations, that caused a shift to the Ising behavior of the real materials. Nevertheless, this
explanation wasn't at all satisfactory, since the critical exponents found in the experiments were completely different from the Ising ones.
In 1992 an elegant theory was proposed by H.Kawamura, who suggested that a finite-temperature phase transition took place in the chiral sector. This scenario
kept also after more modern simulation stated that the model has $T_{SG}>0$. It consists in two steps:
\begin{enumerate}
 \item Spin-Chirality decoupling\\(for a completely isotropic system, as the simulated ones)
 \item Spin-Chirality recoupling\\(for a weakly anisotropic system, as the experimental ones)
\end{enumerate}
The first part claims that the ordering is not simultaneous for both the order parameters (chiral and spin glass overlap). When we decrease temperature from
the paramagnetic phase we first meet at $T_{CG}$ the spontaneous breaking of the discrete $Z_2$ symmetry with the preservation of the continuous $SO(3)$
symmetry, and only at lower temperature $T_{SG}<T_{CG}$ the whole $O(3)$ symmetry is broken. The higher transition, in which only the $Z_2$ symmetry is broken
is called the ``chiral glass transition'', and the phase between $T_{CG}$ and $T_{SG}$, in which only the chiralities are ordered, is called the ``chiral glass state''.

In the second part of this scenario this result in adapted to a real spin glass, with inevitable weak magnetic anisotropy, which reduces the Hamiltonian's
symmetry from $SO(3)\times Z_2$ to only $Z_2$, mixing the chirality to the spin sector. So of the two transitions that take place in an isotropic system,
it would be the chiral glass transition to dominate the real spin glasses' phase transition.

The chirality decoupling scenario explains in a natural way questions on the experimental phase transition like the origin of the non-Ising critical exponents experimentally
observed in canonical spin glasses, and it gave a very elegant explanation of what could be occurring when the early numerical simulations estimated $T_{SG}=0$.
Yet, specially now that it is accepted that $T_{SG}=0$ was a problem of finite size system, or interpolations too far from the critical temperature, it is not a 
dominant scenario in the scientific community, since although there are recent works that would confirm it \cite{kawamura:03}, there are others that give evidence of 
a simultaneous spin and chirality transition \cite{tsgMF3}.

\begin{figure}[!ht]
 \begin{center}
  \includegraphics[width=0.6\textwidth]{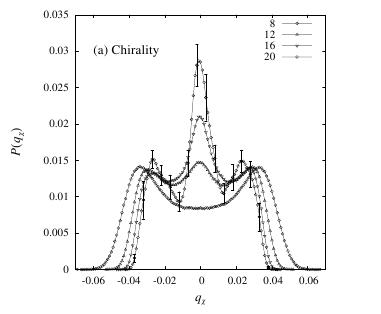}
  \caption{The overlap distribution at $T=0.15$ for the chirality of the $3D$ Heisenberg spin glass with $\pm J$ couplings from \cite[Kawamura (2005)]{pqchir}. This plot suggests that there may be a one-step
RSB for the chirality.}
  \label{fig:P_q-chir}
 \end{center}
\end{figure}

For what concerns the type of ordering, while for the SG order parameter the nature of the glassy phase is debated not differently than with Ising spins, in the chiral sector there has been some
evidence (figure \ref{fig:P_q-chir}) of a one-step RSB \cite{pqchir}, so the $P(q)$ in the thermodynamic limit would be represented by two delta-function peaks at $\pm q_{EA}$ and one at $q=0$.

\subparagraph{Mean Field Analysis}
Evaluating the help that the mean field resolution gives to the understanding of the finite-range interaction model is a troublesome issue. Table \ref{table:MF}, which compares the
mean field critical temperature $T^{MF}_{SG}=\frac{\sqrt{z}}{m}$, where $z$ is the coordination number ($z=6$ in our case), with the estimates of the finite-range $T_{SG}$, shows
us how increasing $m$ it becomes always less reliable. What happens is that the fluctuations neglected with the mean field approximation are more important the larger $m$ is, so
the mean field results with Heisenberg spins are even less reliable than those obtained with Ising spins. We remark also how the data in the table suggests that $T_{SG}=0$ for 
$m\rightarrow\infty$. Unfortunately no mean field solution exists in the chiral sector because it is impossible to define chirality in a non metric space.

\begin{table}[!ht]
\centering 
\begin{tabular}{|c c |c c c|} 
\hline\hline 
$m$ & Model & $T_{SG}^{MF}$ & $T_{SG}$ &  $T_{SG}/T_{SG}^{MF}$\\ [0.5ex] 
\hline 
1 & Ising      & 2.45 & 0.97 \cite{tsgMF1} & 0.40 \\ 
2 & XY         & 1.22 & 0.34 \cite{tsgMF2} & 0.28 \\
3 & Heisenberg & 0.82 & 0.12 \cite{tsgMF3} & 0.15 \\ [1ex] 
\hline 
\end{tabular}
\caption{Comparison of the critical temperature, between mean field approximation and numerical evaluation, varying the number of components of the order parameter. We see that $T_{SG}/T_{SG}^{MF}$
is small and decreases with $m$. Physically this means that fluctuations get larger increasing $m$. Data suggests that $T_{SG}\stackrel{m\rightarrow\infty}{\longrightarrow}0$,
and evidence has been found for this conjecture being true \cite{tsg0}.}
\label{table:MF} 
\end{table}

\section{Dynamics}
At low temperature the dynamics in spin glasses become very slow, such that below the critical temperature the system is always out of equilibrium.
There are many out of equilibrium effects in the dynamics of spin glasses that still haven't been well understood as aging, rejuvenation and temperature cycles.
Yet, in the present work we want to concentrate in another aspect that attracted our curiosity, that is the possibility of a straightforward
analogy between the dynamics of spin and structural glasses.

\subsection{Energy Landscape Scenario}
Although out of equilibrium dynamics is proper of the spin glass phase, an interesting feature could be revealed if we notice that works \cite{picco}
that look for the critical temperature with dynamical criteria tend to overestimate it. This could suggest an analogy with structural glasses,
in which Mode Coupling Theory predicts a power law divergence of the correlation time at $T=T_{MC}>T_c$. This divergence doesn't take place because when 
we get close to $T_{MC}$ different effects come in play, but it could be what is being observed in those works. Goldstein's scenario \cite{goldstein},
which by now is widely accepted, says that the dynamics at low temperature are completely different from those at high temperature. 

When $T$ is low a supercooled liquid explores the phase space mainly by activated jumps between the local minima of the free energy. The characteristic time $\tau$
to jump from one minimum to another is, in agreement with the Arrhenius law $\tau\propto\exp^{-\beta\varDelta V}$, proportional to the exponential of the energy gap.
On the contrary at higher temperatures the typical configurations are not local minima with a positive definite hessian, but saddles with negative directions, 
so the dynamics is much faster. The characteristic time does not represent any more how much it takes for the thermal agitation to cause the barrier jump, but 
how much it takes to the system to individuate the negative directions in the hessian.
We understand, in this scenario, how the decrease of negative directions in lowering the temperature may increase the relaxation time making it apparently diverge, 
just until it becomes of the order of $\tau\propto\exp^{-\beta\varDelta V}$.

Our main aim in this work was to find evidence of the influence of the energy landscape in the dynamics of a spin glass, to give a basis for constructing
a solid connection with the theory of the dynamics of structural glasses. In this perspective we did a deep investigation of the magnitudes that concern
dynamics, such as time correlation functions, and tried to connect them to the energy landscape, by making a massive research of inherent structures (local
energy minima).



\newpage\thispagestyle{empty}\null\newpage 
\chapter{Numerical Techniques}

\minitoc
\mtcskip

\newpage

\section{Thermalization Dynamics}
Monte Carlo (MC) is the most used technique for spin glass computer simulations. It is 
a stochastic Markov Chain method that aims not to sample the whole configuration space, 
but only its most significant parts (importance sampling) in order to extract a fine 
estimate of the observables. 

For Ising spins there exist numerous MC variants that one can easily obtain imposing 
detailed balance to be satisfied, such as \cite{krauth} the most famous Metropolis-Hastings, 
local sampling methods, perfect sampling and heat bath. Al those consist in flipping sequentially
the spins with the constraint of preserving the detailed balance.

For multi-dimensional spins the deal is similar. We will now show the extension of a couple 
of those methods to Heisenberg spins.

\subsection{Metropolis}
\label{Metropolis}
The fingerprint of a Metropolis updating scheme is that new configurations are accepted
with a probability $min[1,e^{-\beta\varDelta E}]$. We generate new configurations by modifying
the orientation of a single spin in the following way:
\begin{description}
 \item A $3-D$ random vector $\vec r$ in the sphere of radius $\delta$ is extracted.
 \item The new spin proposal is
\begin{equation}
 \vec s^{new}_{\vec x} = \frac{s^{old}_{\vec x} + \vec r}
                                 {||s^{old}_{\vec x} + \vec r||}
.
\end{equation}
 \item The new spin is accepted with probability $P=min[1,e^{-\beta\varDelta E}]$ and 
  rejected with probability $1-P$.
\end{description}

We see that the probability of proposing $\vec s^{old}_{\vec x} \rightarrow \vec s^{new}_{\vec x}$
is the same of proposing $\vec s^{new}_{\vec x} \rightarrow \vec s^{old}_{\vec x}$.
The choice of $\delta$ must be sufficiently accurate to give an acceptance rate of $40\%\sim60\%$,
because a too small $\delta$ (hence a too small rejection rate) would make the dynamics too slow,
while a too large rejection rate would keep the system still.

\subsection{Heat Bath}
\label{HB}
The Heat Bath (HB) updating scheme has the main advantage of reducing to zero the rejection 
rate, and for this reason for Heisenberg spins it is much more performing than Metropolis.
The main idea is that, chosen a site, the new spin proposal does not depend by the old 
spin, but it is chosen according to its probability distribution.

For a given site the local field is known, so the conditional probability for the spin's
orientation is
\begin{equation}
 dP(\vec s|\vec h) = d\vec s\frac{e^{\beta J (\vec s\cdot\vec h)}}
                          {\int d\vec s e^{\beta J (\vec s\cdot\vec h)}}
.
\end{equation}
We can make a wise reference change to coordinates $(x',y',z')$ with $z'$ $//$ $\vec h$ to 
get trivial scalar products, and then pass to polar coordinates, with $s=(\cos\theta'\cos\phi',
\sin\theta'\cos\phi',\sin\phi'=z')$:
\begin{equation}
 dP(\vec s|\vec h) = d\phi d\theta\frac{e^{\beta J h \sin\theta} \sin\theta}
                     {\int d\phi d\theta      e^{\beta J h \sin\theta} \sin\theta}
.
\end{equation}
This way we get a factorized probability distribution function
 \begin{equation}
 dP(\vec s|\vec h) = \frac{d\phi }{2\pi} \times dz'
  \frac{e^{\beta J h z'}}
  {\int dz' e^{\beta J h z'}}
 \end{equation}
where $h=||\vec h||$. Since the angle $0<\phi'<2\pi$ is uniformly distributed it's easy to
get its pseudo-random value. For $z'$ we use $u=e^{\beta J h z'}$ that is uniformly 
distributed in the interval $e^{-\beta J h}<u<e^{\beta J h}$.

At this point we have to come back to our original frame with minimum computational effort.
To do so we remark that our spin is
\begin{equation}
 \vec s = z' \frac{\vec h}{h} + \sqrt{1-{z'}^2}\vec R_\bot
\end{equation}
where $\vec R_\bot$ is a unitary random vector, orthogonal to the local field $\vec h$,
that we can obtain by extracting a random unitary vector, peeling its component parallel
to $\vec h$, and renormalizing it.

\subsection{Over Relaxation}
\label{OR}
The present is a deterministic, very computationally-fast, microcanonical simulation scheme. Although its
dynamics are not physical, it helps when we want to sweep quickly the configuration space.
Performing $1$ HB sweep every $L$ Over Relaxation (OR) sweeps we generate in the lattice
an iso-energetic spin wave after each energy change. Of course it cannot reproduce real
dynamics, but it is very useful to generate thermalized configurations of the system.

The algorithm proposes the maximum spin change that leaves the energy invariant so the
new spin will always be accepted (see the Metropolis acceptance rate in section \ref{Metropolis}).
The new spin is so binded the condition
\begin{equation}
 \vec s^{new} \cdot \vec h = \vec s^{old} \cdot \vec h 
\end{equation}
and we obtain the maximal change by reversing the spin's component that is perpendicular 
to the local field $\vec h$:
\begin{equation}
 \vec s^{new} = 2\vec h_{\vec x}\frac{\vec h_{\vec x}\cdot\vec s_{\vec x}}{\vec h_{\vec x}^2}-\vec s_{\vec x}
\end{equation}

\section{Minimizing the Energy}
Obtaining spin glass low energy configurations is computationally hard\cite{NP}, and
this makes our objective of finding inherent structures a real challenge. While for 
Ising spins there exists a whole literature of numerical methods used to minimize the energy
such as simulated annealing \cite{SA}, cluster algorithms \cite{hartmann:02}, genetic
algorithms\cite{GA,GAbook}, and combinations of those methods\cite{pal,hartmann:01},
almost nothing exists for Heisenberg spin glasses.

We will now give a review of the small amount of methods used for Heisenberg spin 
glasses, and present our choice, which is not applicable to Ising spins and is new for
HSG.

\subsection{Gauss-Seidel}
\label{subsec:gauss-seidel}
The most trivial and common way to find low-energy states is performing
a quench.

It is easy to see that the energy of a single site
\begin{equation}
 e_{\vec x} = -\vec s_{\vec x} \cdot \vec h_{\vec x}
\end{equation}
is minimized if the spin $\vec{s}_x$ is parallel to the local field $\vec{h}_x$.
this observation can suggest us a quite brutal way to lower the energy, i.e. aligning
sequentially the spins to the local field, i.e. making the substitution
\begin{equation}
  \vec s_{\vec x} \rightarrow \frac{\vec h_{\vec x}}{||\vec h_{\vec x}||}
\end{equation}
This type of method is much older than the mere existence of spin glass models. In fact
we will see that it is exactly the application to our model of the Gauss-Seidel method,
used to solve systems of linear equations\cite{sokal}. In fact the problem solving the 
linear system $A\phi=f$, where $A$ is a given symmetric positive-definite matrix and 
$f$ is a given vector, is equivalent to the one of minimizing the Hamiltonian
\begin{equation}
 H(\phi)=\frac{1}{2}(\phi,A\phi)-(f,\phi).
\label{eq:quadratic}
\end{equation}
Given the Hamiltonian in equation \ref{eq:quadratic}, the Gauss-Seidel Algorithm consists 
in \emph{sweeping through the sites $i$, and at each 
step replacing $\phi_i$ by that new value $\phi_i'$ which minimizes $H$ when the other
fields $\{\phi_j\}_{j\neq i}$ are held fixed at their current values}\cite{sokal}. 
In other words, if we write
\begin{equation}
 H(\phi_i,{\phi_j}j\neq i) = \frac{1}{2}A_{ii}(\phi_i-b_i)^2 + c_i
\end{equation}
the Gauss-Seidel algorithm consists in simply replacing $\phi_i$ with $b_i$.
This is exactly what we do with the quench routine when we substitute $\vec s_{\vec x}$ 
with $\frac{\vec{h_{\vec x}}}{||\vec{h_{\vec x}}||}$, so if we show that the 
EA-Heisenberg Hamiltonian is reducible to the same form, we can state immediately this
algorithm's main property: it's ineffectiveness, and seek for a better solution. As a 
matter of fact this algorithm has an extremely slow rate of convergence \cite{varga}. 
For example for $A=\Delta + m^2$ the convergence time grows as $m^{-2}$.
 
\paragraph{Reduction of the EA Heisenberg Hamiltonian to a quadratic form}

Let's show how, when we are close to a local minimum, our Hamiltonian can be reduced to
a quadratic form in chiral perturbation theory.

If we call $\{\vec s^0\}$ a generic inherent structure configuration, i.e. with all
the spins aligned to the local field, any spin is expressible as
\begin{equation}
 \vec s_{\vec x} = \vec s^0 \sqrt{1-\vec\pi_{\vec x}^2} + \vec \pi_{\vec x}
\end{equation}
with the condition
\begin{equation}
\label{perpendicular}
 \vec h_{\vec x} \cdot \vec\pi_{\vec x} =  \vec s^0_{\vec x} \cdot \vec\pi_{\vec x} = 0.
\end{equation}
$\vec\pi$ is the component of the spin perpendicular to the local field, and we call it 
pion. If $|\vec\pi|$ is small, i.e. we are close to the local minimum, we can develop 
the Hamiltonian to the second order in $\vec\pi$, keeping in mind condition 
\ref{perpendicular},  in order to obtain
\begin{equation}
 H(\{\vec s\}) = H(\{\vec s^0\}) 
 + \frac{1}{2}\displaystyle\sum_{\vec x}(\vec h_{\vec x} \cdot \vec s^0_{\vec x})\vec\pi^2_{\vec x}
 + \displaystyle\sum_{\vec x}\vec\pi_{\vec x}\cdot(\sum_{||\vec x- \vec y||=1}J_{\vec x\vec y}\vec\pi_{\vec y})
.
\end{equation}
This way the variation $\varDelta H$ can be written as a quadratic form in an $N$-dimensional space:
\begin{equation}
 \varDelta H = H(\{\vec s\}) - H(\{\vec s^0\}) = -\frac{1}{2}(\{\vec\pi\},M\{\vec\pi\}) + o(\vec\pi^3)
\end{equation}
with
\begin{equation}
M_{xy} = \delta_{\vec x\vec y}-\displaystyle\sum_{\mu=\pm1}^{\pm D}\delta_{\vec y\vec x+\vec e_\mu} J_{\vec x\vec y}
,
\end{equation}
so finding an inherent structure is actually equivalent to minimizing a quadratic form
when we're in the limit of small pions.

\subsection{Genetic Algorithms}
Other than brutally applying the Gauss-Seidel method, very little has been done for 
Heisenberg spin glasses. Nevertheless, in a recent article \cite{giappa} from Y.Iyama 
and F.Matsubara we can see an attempt to extend the genetic algorithm thought up by 
P\'al \cite{pal} for Ising spins, to Heisenberg
spins, with the intention of finding ground states. Since this is the only other kind
of energy-minimizing routine we found in literature we will explain the ideas on which 
it is based.

The basic idea of a genetic algorithm is to find a low energy state mixing in a smart
way the parts of a selection parent configurations (from analogy with reproduction the
name genetic, since it is as if the configurations pass to their heirs their genetic 
code). Clearly the initial choice of the parent configurations is a crucial step. 

P\'al's suggestion \cite{pal} is to use a hybrid algorithm which alternates the genetic 
algorithm steps with a generic non-genetic minimization algorithm, and this is what is done by Iyama and 
Matsubara in \cite{giappa}, who chose to alternate quenches (Gauss-Seidel) to the genetic steps.

It appears, from their work, that a very big computational effort is necessary to find
low energy configurations with their algorithm, since they study very small lattices
($L\leq13$), and the number of samples studied is not high ($500$ samples for $L=13$).

Later in this paper we will refer to this work since it is the only example in literature of search
of ground states in Heisenberg spin glasses (although they study $\pm J$ bonds, while 
ours have a Gaussian distribution), and they give the only existing (to now) estimate of the 
ground state energies. In a brief study of $\pm 1$ bonds we will show how
results comparable to theirs are achievable with much smaller computational effort.

\subsection{Successive Over Relaxation}
It is common sense that greedy algorithms such as Gauss-Seidel are not very effective,
especially compared to ungreedier versions of their selves. Taking advantage of this 
precious advice we can improve this scheme in a simple and effective way. Instead of 
choosing the $\phi_i$ that minimizes the Hamiltonian we can make an apparent non-optimal choice
that lowers the energy, but less. On the long run the convergence time is definitely 
quicker (for $A=\Delta + m^2$ the convergence time grows as $m^{-1}$ instead of 
$m^{-2}$ with the best choice of the parameters).

In our problem, if we call
\begin{equation}
 \vec s^{GS}_{\vec x} = \frac{\vec h_{\vec x}}{||\vec h_{\vec x}||}
\end{equation}
the new spin that minimizes the local Hamiltonian, and
\begin{equation}
 \vec s^{OR}_{\vec x} = 2\vec h_{\vec x}\frac{\vec h_{\vec x}\cdot\vec s_{\vec x}}{\vec h_{\vec x}^2}-\vec s_{\vec x}
\end{equation}
the spin deriving from the Over Relaxation algorithm described in section \ref{OR}, the 
Successive Over Relaxation is expressed by choosing the new spin as
\begin{equation} 
\vec s^{SOR}_{\lambda,\vec x} = \frac{\vec s^{GS}_{\vec x} + \lambda \vec s^{OR}_{\vec x}}
                                   {||\vec s^{GS}_{\vec x} + \lambda \vec s^{OR}_{\vec x}||}
\end{equation}
where $\lambda>0$ is the parameter that determines how ungreedy the routine is. This way we
are choosing our new spin damping the maximum energy minimization with the iso-energetic 
change that most varies the Hamiltonian.

\begin{figure}[!ht]
 \includegraphics[width=\textwidth]{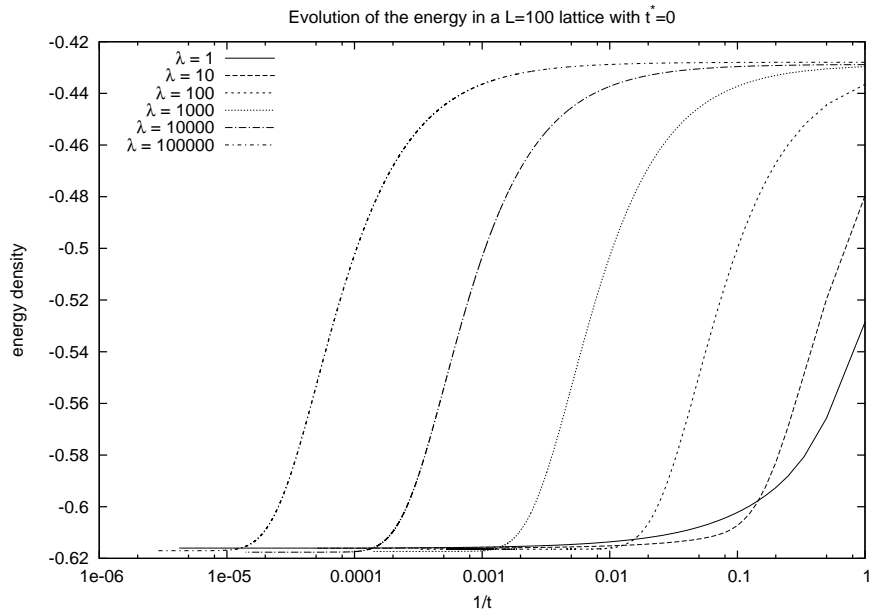}
\caption{Successive Over Relaxation in a $L=100$ after having performed $t^*=0$ Gauss-Seidel steps before starting with SOR.
With small $\lambda$ the dynamics is similar to a quench: the system dives into a valley of the free energy without selection,
and it stays there a lot of time before converging, as we know that with the Gauss-Seidel method convergence is a bad issue.
On the other side dynamics with very large $\lambda$ spend a lot of time at high energy: they select better the valley where
to get inside, so they reach lower energies (it will be more evident in the zoomed figure \ref{fig:tstarperde}) and once they're
in a valley they converge faster. We can remark the too greedy routines, which converge too slowly because they spend too much 
time at low energies, and the too greedy ones, that stay too much time at high energies.}
\label{fig:en_evol}
\end{figure}

Figure \ref{fig:en_evol} shows how the energy decrease is initially lower with higher 
$\lambda$'s, but there is a time after which the ungreedier routine (greater $\lambda$) 
becomes more efficient.

We remark the two trivial limits in $\lambda$ for this algorithm:
\begin{eqnarray}
 \displaystyle\lim_{\lambda\rightarrow0} s^{SOR}_{\lambda}     & = \vec s^{GS}\\
 \displaystyle\lim_{\lambda\rightarrow\infty} s^{SOR}_{\lambda}& = \vec s^{OR}
.
\end{eqnarray}
It is clear that $\lambda$ has to be chosen with care, because a small one risks to be
too greedy, hence too similar to a brutal quench, while a too big one may be too 
ungreedy, having a too long plateau, and never converge.

\paragraph{Variants}
If we want our routine to be more site-dependent we can slightly modify it as follows:
\begin{equation}
\vec s^{new}_{\lambda,\vec x} = \frac{\vec h_{\vec x} + \lambda \vec s^{OR}_{\vec x}}
                                   {||\vec h_{\vec x} + \lambda \vec s^{OR}_{\vec x}||}
.
\end{equation}
This way the dynamics will be less damped in the sites where the local field is strong,
and more where it is weak.
This is the variant we used throughout our investigation.

Another idea could be to alternate $aL$ ($a=1,2$) Over Relaxation steps with one quench
steps instead of updating the spins with always in the same ungreedy way. This would mean
that before lowering the energy we sweep the system with an iso-energetic wave of spins.
We tested this method both alternating GS and OR, and SOR and OR. The result was more 
satisfying than pure Gauss-Seidel, but the convergence time resulted being less convenient 
with respect to Successive Over Relaxation.

\section{Computing Observables}
A main advantage of simulations is that we can know the exactly properties of each 
single particle of the system, while in real experiments only some (global) observables
can be found by analysing the reaction to determinate stimuli. The possibility of
computing observables directly represents a great advantage and it pays off the 
problem of small sizes.

The observables that we computed in our simulations were:
\begin{itemize}

 \item Energy
 \item Overlaps
 \item Correlation Lengths and Functions
 \item Magnetic Susceptibility
 \item Energy-Overlap Correlation
 \item Binder Cumulants
 \item Correlation Times and Functions

\end{itemize}
Let's analyse them one by one.

\subsection{Energy}
The energy is the most intuitive and basic observable to look for in a simulation, and
it represents one of the great challenges of SG systems, since due to the frustration
it is very hard to find low energy configurations, and in fact this is what we're going 
for.

It can be found very trivially, as it is defined by the model's Hamiltonian:
\begin{equation}
 H = \frac{1}{2} \displaystyle\sum_{\vec x} \vec s_{\vec x} \cdot \vec h_{\vec x}
\end{equation}
It simply requires a sweep over all the spins summing the energy per site:
\begin{equation}
 e_{\vec x} = \frac{1}{2} \vec s_{\vec x} \cdot \vec h_{\vec x}
\end{equation}

\subsection{Overlaps}
There are two types of overlap to compute, the CG overlap, and the SG. We used the overlap definitions anticipated in 
chapter \ref{chap:one}. The CG overlap was the scalar quantity
\begin{equation}
 E_{CG} = \frac{1}{3}\displaystyle\sum_{\mu=1}^3 E^\mu = \frac{1}{3N}\sum_{\mu=1}^3\sum_{\vec x}\epsilon^\mu(\vec x)
,
\end{equation}
while for the SG overlap we used a slightly more complicated object, also introduced in chapter \ref{chap:one}, that
guarantees rotational invariance. As a matter of fact it is not an overlap, but a square overlap:
\begin{equation}
  q^2 = \frac{1}{N^2} tr (Q Q^\dagger)
.
\end{equation}

\paragraph{Properties of $q^2$}
Let's see the typical values of $q^2$. We remind that the tensors $Q_{\alpha\beta}$ are defined as
\begin{equation}
 Q_{\alpha\beta} = \displaystyle\sum_{\vec x} \tau_{\alpha\beta}(\vec x) = \sum_{\vec x} s_{\vec x, \alpha}^{(a)}s_{\vec x, \beta}^{(b)}
,
\end{equation}
so, by substituting this definition and rearranging the sums we get, with trivial calculation,
\begin{equation}
\label{eq:ovlp}
 tr (Q Q^\dagger) = \displaystyle\sum_{\vec x}\sum_{\vec y}(\vec s_{\vec x}^{(a)}\cdot\vec s_{\vec y}^{(a)})
                                                           (\vec s_{\vec x}^{(b)}\cdot\vec s_{\vec y}^{(b)})
,
\end{equation}
that may be easily bounded thanks to the Cauchy-Schwarz inequality, that guarantees us that $\vec s_{\vec x}\vec s_{\vec y} \leq \sqrt{\vec s_{\vec x}^2
\vec s_{\vec y}^2}=1$, so
\begin{equation}
tr (Q Q^\dagger) \leq N^2
,
\end{equation}
and the square overlap, with the normalization that we chose, gets the bound
\begin{equation}
 q^2 = \leq 1
\end{equation}

\subparagraph{SG Self-overlap}
We want to show the typical value of the SG self-overlap for a random configuration. To evaluate it we impose $(a)=(b)$ in eq. \ref{eq:ovlp}, 
and obtain
\begin{eqnarray}
 tr (Q_{self} Q_{self}^\dagger) = & \displaystyle\sum_{\vec x}\sum_{\vec y}(\vec s_{\vec x}\cdot\vec s_{\vec y})^2 \\
				= & \displaystyle\sum_{\vec x}\vec s_{\vec x}^4 + \sum_{\vec x\neq \vec y}(\vec s_{\vec x}\cdot\vec s_{\vec y})^2
.
\end{eqnarray}
 To estimate the average of the product $(\vec s_{\vec x}\cdot\vec s_{\vec y})^2$ we can put ourselves in a reference frame $(\hat{x},\hat{y},\hat{z})$ with 
$\vec s_{\vec x} // \hat{z}$. The value of the scalar product is then given by the third component of $\vec s_{\vec y}$, $s_{\vec y,z}$. This way
\begin{equation}
 (\vec s_{\vec x}\cdot\vec s_{\vec y})^2 = s_{\vec y,z}^2
\end{equation}
is a uniformly distributed random variable whose mean value is $1/3$, since our system is completely isotropic.

We can now determine the average self-overlap
\begin{eqnarray}
 <q^2_{self}> = & \frac{1}{N^2}(N+\displaystyle\sum_{\vec x\neq\vec y}<(\vec s_{\vec x}\cdot\vec s_{\vec y})^2>)\\
	      = & \frac{1}{N} + \frac{1}{3} - \frac{1}{3N} \stackrel{N\rightarrow\infty}{\longrightarrow}\frac{1}{3}
\end{eqnarray}


\subsection{Correlation Lengths and Functions}
\label{subsec:corr}
An extremely important quantity for our simulations was the correlation length, because
not only it gives information regarding the properties of a state (e.g. it is zero at 
infinite temperature while it diverges at the phase transition), but since it estimates
until which distance two spins can be considered correlated, it indicates us if we can 
consider the system in thermodynamic limit or not. For the latter check we can also see
if the correlation functions go to zero within half lattice size
\footnote{Since the boundary conditions are periodic the maximum distance between two
points is $L/2$, not $L$.}.

The correlation function relative to a generic observable $\phi$ is defined as:
\begin{equation}
 G(\vec{r}) = \displaystyle\sum_{\vec{x}} \phi_{\vec{x}}\phi_{\vec{x}+\vec{r}}  
\end{equation}

As it is shown in appendix \ref{app:fourier}, together with a computationally smart way to write it, 
a way to compute the correlation length 
$\xi$ involves the correlation function's Fourier transform $\hat{G}(\vec k)$:
\begin{equation}
\label{eq:xi-fourier}
 {\xi_F}^2 = \frac{1}{4 sin^2 (\frac{\pi}{L})} [\frac{\hat{G}(\vec 0)}{\hat{G}(\vec k_{min})} - 1] = 
             \frac{1}{4 sin^2 (\frac{\pi}{L})} [\frac{\chi}{F} - 1]
\end{equation}
where we stressed that the magnetic susceptibility $\chi$ is the $\vec k=0$ Fourier transform
of the correlation function:
\begin{eqnarray}
 \chi = \hat{G}(\vec 0) \\
    F = \hat{G}(\vec k_{min})
\end{eqnarray}
and due to the periodic boundary conditions $\vec k_{min}= (\frac{2 \pi}{L},0,0)$.

\paragraph{Integral Estimators}

If we define an integral estimator $I_k$ as:
\begin{equation}
 I_k = \displaystyle\int^{\frac{L}{2}}_0 d\vec{r} \vec r^k G(\vec r)
\end{equation}
we can readily see that (appendix \ref{app:estimators}) in the thermodynamic limit we 
can define two more ways to find the correlation length in a $D$-dimensional system:

\begin{equation}
\xi^{(02)} = \sqrt{\frac{I_{D+1}}{I_{D-1}}} \simeq \xi
\end{equation}
\begin{equation}
 \xi_{k,k+1} = \frac{I_{k+1}}{I_k} \varpropto \xi
\end{equation}

\begin{figure}[!ht]
\begin{center}
  \includegraphics[width=\textwidth]{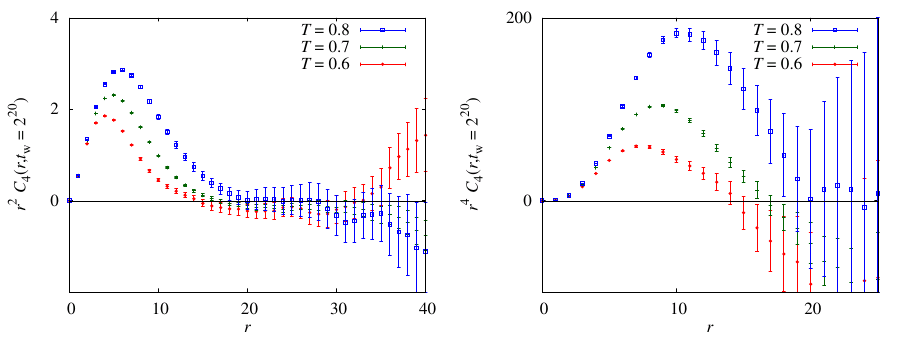}
  \caption{Comparison between $\vec r^2 G(r)$ and $\vec r^4 G(r)$ from simulations 
of an $L=80$ Ising spin lattices at three different temperatures. In the first case 
the significant data is clearly out of the noise-dominated region, while in the second
the maximum is pushed rightwards giving plenty of noise to the calculation of $I_4$.
Figure borrowed from \cite{corr:01}.}
  \label{fig:estimators}
\end{center}
\end{figure}

All these definitions suffer from systematic errors because they are derived from the 
long distance decay of $G$:
\begin{equation}
 G(\vec r) \sim \frac{1}{r^a}\exp{(-\frac{r}{\xi})}
\end{equation}
that is only an asymptotic formula for large values of $r$. The systematic errors in 
these definitions can be reduced by considering a large value of $k$ (since the $r^k$ 
factor would suppress the deviations at short distances). However, there is also 
the issue of statistical errors to consider. As it can be seen in figure 
\ref{fig:estimators}, borrowed from reference \cite{corr:01}, a large value of $k$ 
pushes the maximum of $\vec r^k G(\vec r)$ towards a zone where the statistical error
becomes too big. Therefore it is necessary to find a compromise $k$.

\paragraph{Plane to plane correlation functions}

We can define three types of plane to plane correlation functions, one for each direction. To avoid
making notation heavy we write it explicitly for a given direction: 
\begin{equation}
C^x(r) = \frac{1}{N}\displaystyle\sum_{x=0}^{L-1} P(x) P(x+r)
\end{equation}
\begin{equation}
P(x) = \displaystyle\sum_{y,z} \phi_{x y z} 
\end{equation}
where $\phi_{x y z}$ is any observable with which a correlation function has sense.
Since the space is perfectly isotropic we can reduce the error by calculating
\begin{equation}
 C(r)=\frac{1}{3}(C^x(r) + C^y(r) + C^z(r))
.
\end{equation}

Plane to plane correlation functions represent a fine way to have more precise estimates 
of the correlation length $\xi$, since beyond being a natural choice for square lattices, 
in which we can easily average it on the euclidean directions, and they decay far more 
slowly than the point to point correlation functions.

In fact if we define the plane to plane integral estimators:
\begin{equation}
  J_k = \displaystyle\int^{\frac{L}{2}}_0 dr r^k C(r)
\end{equation}
where we called $C(r)$ the mean of the plane to plane correlation functions in all the
directions, it is immediate to see that since the space is homogeneous and isotropic
\begin{eqnarray}
 J_k \sim 4\pi I_{k+2} \\
  \chi= 2 J_0 -C(0)-C(L/2)
\end{eqnarray}
so for the same integral estimator we need to use lower powers of $r$ if we're using 
the $J_k$'s. This means that $r^k C(r)$ will have sense in a 
larger span of distances, so the amount of data without signal will drastically 
decrease.


\subparagraph{Truncating data}

Another thing that can be done to reduce consistently the statistical error, in favor
of a small bias, is to do as proposed by Martín-Mayor,Marinari et al. in \cite{corr:01},
 i.e. to use a self-consistent integration cut-off as it has always been done with the
 study of correlated time series \cite{amit}.

What we do is to decide \emph{a posteriori} a critical distance $r_c$ beyond which 
collecting data has no sense because the measure is dominated by the statistical 
error. This distance was chosen to be the first one at which the error became greater
than one third of the correlation function:

\begin{equation}
  I_k = \displaystyle\sum^{\frac{L}{2}}_0 \vec r^k G(\vec r) \longrightarrow 
  I_k = \displaystyle\sum^{r_c}_0 \vec r^k G(\vec r)
\end{equation}
\begin{equation}
 r_c \equiv \{first\ \ r : C(r+1)<3\sigma(r+1)\}
\end{equation}
A truncation of this type in our work has reduced the error of several orders of 
magnitude, giving a quite small bias \cite{corr:01} (one or two percents), that could be seen from the change 
of the value of the magnetic susceptibility $\chi$.

\subsection{Magnetic Susceptibility}
In theoretical physics, the magnetic susceptibility is defined as the zero-moment Fourier transform of the correlation function
\begin{equation}
 \chi = \hat{C}(k=0) = \displaystyle\sum_r C(r)
\end{equation}

The magnetic susceptibility that we calculate in our simulations is what in the experimental jargon is 
called the non-linear susceptibility. If we put a spin glass under a uniform magnetic field $h$, the magnetization $m$ can be written as a series
of $h$
\begin{equation}
 m(h) = \chi_L h + \chi_{NL} h^3 + ...
\end{equation}
The term $\chi_L$ is the linear magnetic susceptibility $\frac{\partial m}{\partial h}$ that does not diverge at $T_{SG}$. What diverges is the non-linear magnetic
susceptibility $\chi_{NL}$ and that's what makes it a more interesting observable.


\subsection{Binder's Cumulants}
\label{subsec:binder}
Binder's cumulants are ratios of cumulants of the order parameter that give us a good 
comprehension of the finite-size effects. They are often used to individuate phase transitions
because at $T_c$ they are size-invariant. In our case we used them to have a check of
how close to the thermodynamic limit we were throughout our work.

Due to the presence of two order parameter, we can define two Binder Cumulants. The chiral-glass
Binder Cumulant is
\begin{equation}
 B_L^{CG} = \frac{<E_{CG}^2>^2}{<E_{CG}^4>}
\end{equation}
while the spin glass cumulant is
\begin{equation}
 B_L^{SG} = \frac{<tr(QQ^\dagger)>^2}{<tr(QQ^\dagger)^2>} = \frac{<q^2>^2}{<q^4>}
.
\end{equation}

It is possible to show that the asymptotic values of these cumulants for systems at infinite temperature in the thermodynamical
limit are
\begin{itemize}
 \item $B_\infty^{SG}=11/9$
 \item $B_\infty^{CG}=3$
 \end{itemize}
The interested reader can refer to appendix \ref{app:binder} for the calculation of $B_\infty^{SG}$.

\subsection{Self-correlation Times and Functions}
\label{sec:corrtime}
The self-correlation time $\tau$ represents the amount of time in which a system loses all the memory of its initial configuration. It is calculated
from the time-correlation functions $C(t)$, that give us an idea of how much the system at time $t$ is correlated with itself at time $0$.
The time-correlation functions measure how much a particular observable changes during time, so it is evident that it s possible to define
a whole set of self-correlation times and functions, and that the most important of the times is the greatest, since it is the characteristic time
after which \emph{all} the observables have no correlation with their initial values.

In our work these quantities have been used both to know how much we needed to wait for the thermalization of the system, and to measure the scaling
of the $\tau$ in the paramagnetic phase.

We used three self-correlation functions:
\begin{itemize}
 \item Scalar Correlation function
\begin{equation}
 C_{Scalar}(t) = < \vec s_{\vec x}(0) \cdot\vec s_{\vec x}(t) > = \frac{1}{N}\displaystyle\sum_{\vec x} \vec s_{\vec x}(0)\cdot \vec s_{\vec x}(t)
\end{equation}

 \item Tensorial Correlation function
\begin{equation}
 C_{Tensorial}(t) = \frac{1}{N^2}tr(Q_{0,t}Q_{0,t}^\dagger)
\end{equation}
where $Q_{0,t}$ is the overlap calculated between configurations of the same replica at times $0$ and $t$.

 \item CG Correlation function
\begin{equation}
 C_{Scalar}(t) = \frac{1}{3N}\displaystyle\sum_\mu\sum_{\vec x} \kappa^\mu_{\vec x}(0) \kappa^\mu_{\vec x}(t)
\end{equation}

\end{itemize}

And calculated the correlation times:

\begin{eqnarray}
 \tau_{Scalar} &=& \displaystyle\int_0^\infty C_{Scalar}(t)dt\\
 \tau_{CG}     &=& \displaystyle\int_0^\infty C_{CG}(t)dt
\end{eqnarray}
Just as in the case of the spatial correlation times, we truncated the data when $C(t)$ was smaller that $3$ times its error bar. We did not calculate
$\tau_{Tensorial}$ because the square overlap does not go to zero for uncorrelated samples, but to $1/N$, so the data would be biased and not 
$L$-independent, especially when truncating data.

\paragraph{Aging}
It looks obvious that if one measures the self-correlation function starting not from an initial time $0$, but from a time $t_0$, 
nothing should change. Nevertheless, in spin glasses this time-translational invariance is not so obvious. In fact what happens is that while this
symmetry is verified at high temperatures, below $T_{SG}$ the value of the self-correlation function depends on the amount of time we wait before 
starting our measures. This phenomenon, called aging, obliges us to redefine the correlation functions as $C(t_w,t)$, where $t_w$ is the time we 
wait before starting to collecting data.

\begin{figure}[h]
 \includegraphics[height=0.4\textheight]{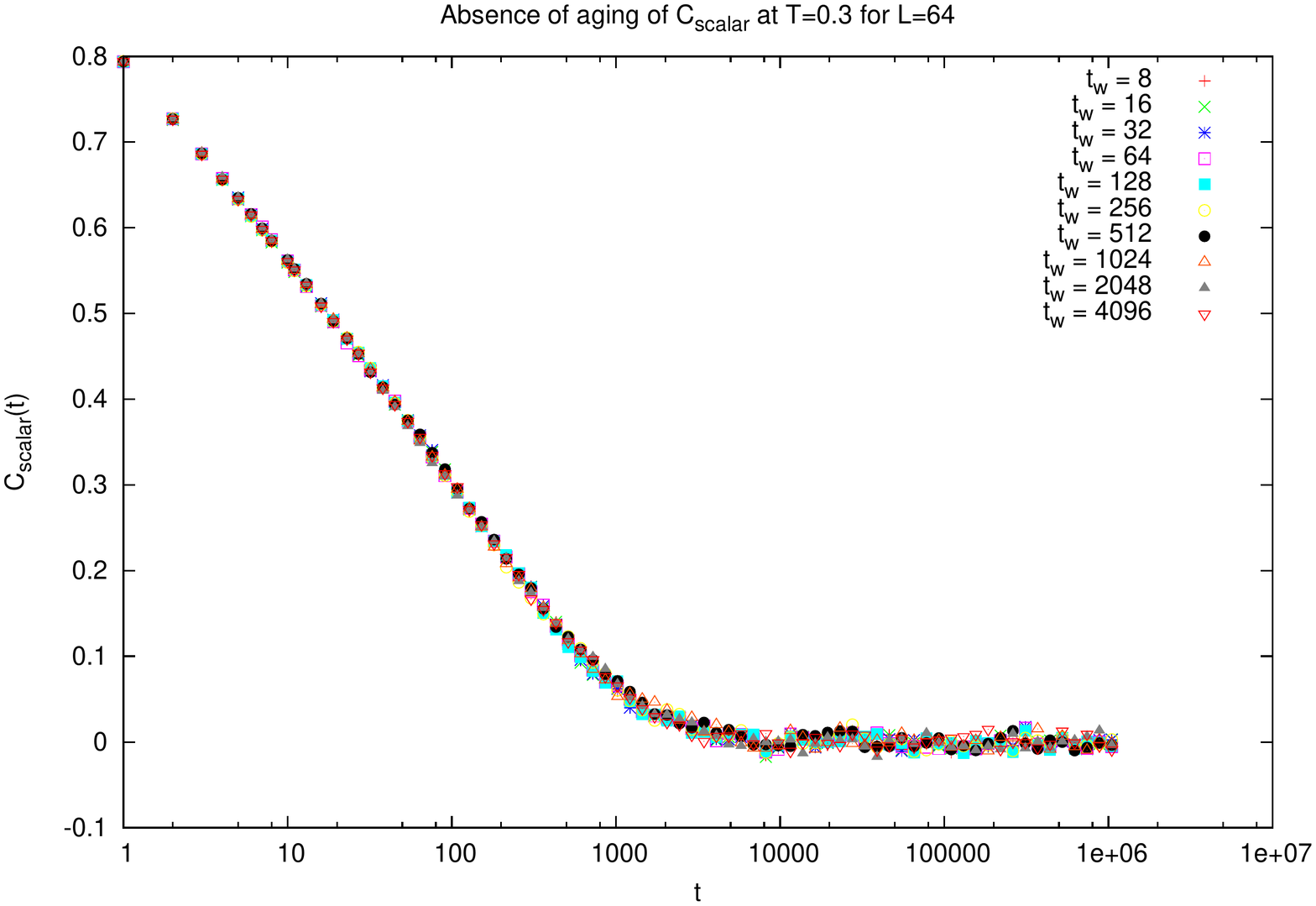}
 \includegraphics[height=0.4\textheight]{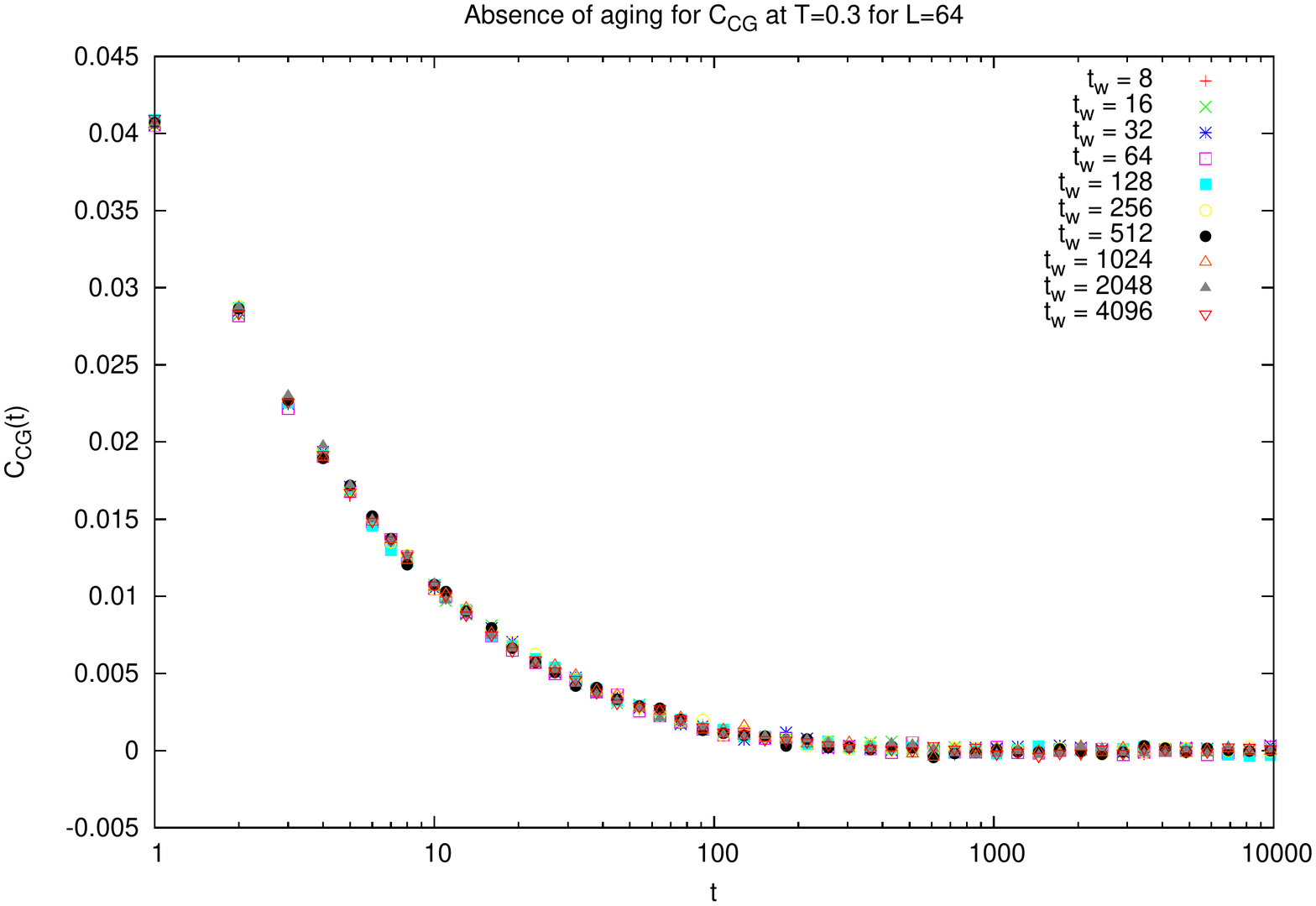}

 \caption{The scalar and chiral-glass time-correlation function $C_{scalar}(t_w,t)$ and $C_{CG}(t_w,t)$ in an $L=64$ lattice at $T=0.3$. 
 As we can see there is no aging in the paramagnetic phase, since data for different $t_w$'s collapse perfectly.}
\label{C_escalar_inv_T030_L64}
\end{figure}

We measured self-correlation functions from many $t_w$, to verify that we did not have aging. Indeed the data was independent from $t_w$ as we can
see from the collapse of the data in figures \ref{C_escalar_inv_T030_L64}
where we plotted $C(t;t_w)$ for many $t_w$. It is because $C(t_w,t) = C(t-t_w)$ that in this report we refer to it as $C(t)$.

%

\chapter{Results}

\minitoc
\mtcskip

\newpage

\section{Using Gauss-Seidel}
Although we showed in chapter \ref{subsec:gauss-seidel} that the Gauss-Seidel algorithm is not effective for
our problem, a complete analysis should start from the simplest cases to be able to 
make comparisons which make us understand if different routines really do improve the
dynamics or not.

\begin{figure}[h]
 \includegraphics{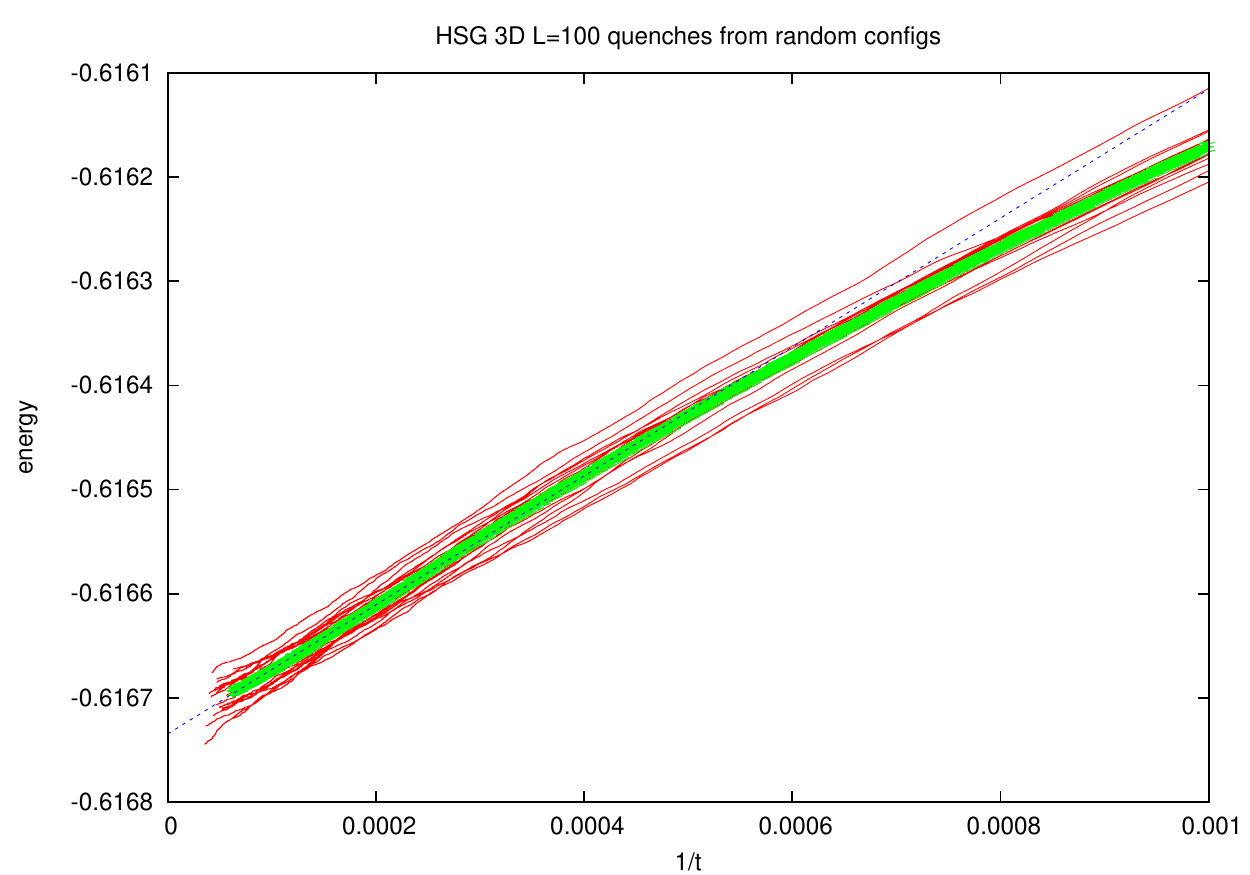}
 \caption{Evolution of the energy density during a quench in a $L=100$ lattice. 
  The plot shows $20$ quenches from random configurations (red lines), i.e. from 
  $T=\infty$ configurations, which were stopped when the energy gain after a sweep 
  was less than $10^{-10}$. The green line is the mean with error bars, which was 
  performed only in the points where all the data was present, since the duration
  of the runs was not even. The dashed blue line is a fit in the linear regime
  $t^{-1}<0.0005$, which yields a good $\chi^2$ test.}
\label{fig:longquench}
\end{figure}

In this perspective we investigated the convergence of the Gauss-Seidel algorithm in 
quite large lattices, and as it is shown in figure \ref{fig:longquench} the convergence
rate is with a power law, which is far too slow for our needs.

At the beginning of our work our convergence criterion has been checking the energy gain
in the last sweep, but we readily realized that a more solid one was to see the maximum
pion in the lattice, since in an inherent structure the pion field is zero.

\section{Using Successive Over Relaxation}
A disadvantage of Successive Over Relaxation respect to classical quenches is that it 
depends from the parameter $\lambda$, therefore it's performance is not univocally 
determined, and it is necessary to do some preliminary studies to choose the optimal
one. We demand a good algorithm the capability of converging to low energies in an 
affordable amount of time.

Since at the beginning of a simulation a greedy routine such as Gauss-Seidel descends
in energy much faster than an ungreedy one such as Successive Over Relaxation, one could
argue that to increase the convergence speed it could be convenient to start with the
Successive Over Relaxation only after having performed $t^*$ quench steps.
This increases the number of non-physical parameters of our routine to two: $\lambda$
and $t^*$.

\subsection{Parameter choice}
To decide how to choose $t^*$ and $\lambda$ we ran simulations on a large lattice 
seeking for the best performances.

\begin{figure}[!ht]
 \includegraphics[width=\textwidth]{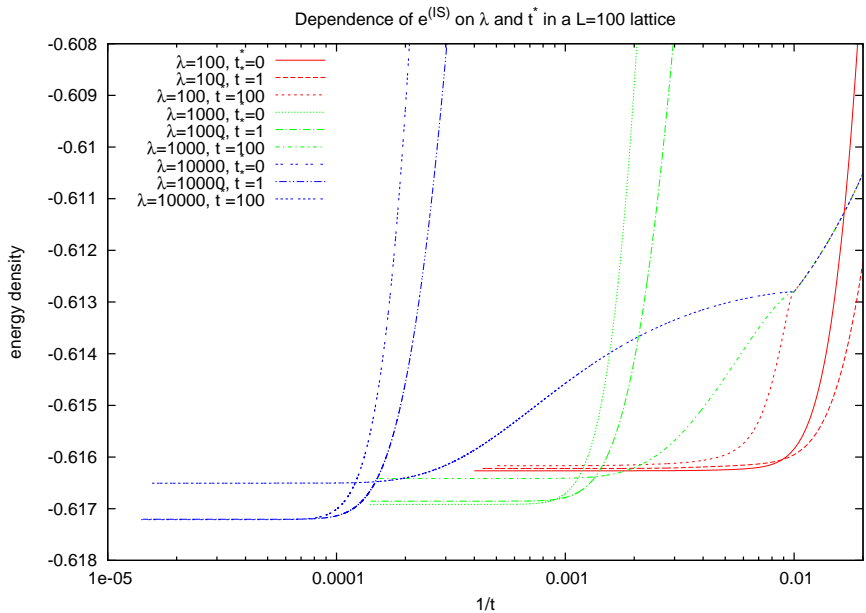}
\caption{The evolution of the energy for $3$ different $\lambda$'s, and $t^*$'s in a $L=100$ lattice. Each $\lambda$ has an associated color. The runs stopped
when the maximum spin component perpendicular to the local field was smaller than $10^{-10}$. It is possible to see the point $t^*=100$ where the 
Gauss-Seidel routine stops in favor of the SOR. Depending on $\lambda$, after this point, the evolution at this point is faster or slower than the 
quench. The most ungreedy $\lambda$ on the long run reaches a lower energy, but at the cost of a much longer convergence time.
What appears clear is that the convergence time depends exclusively on $\lambda$, and also that the greater the quantity of quench steps before 
turning to SOR, the higher are the energies achieved. Hence, there is no point on having $t^*>0$.}
\label{fig:tstarperde}
\end{figure}

\begin{figure}[!ht]
 \includegraphics[width=\textwidth]{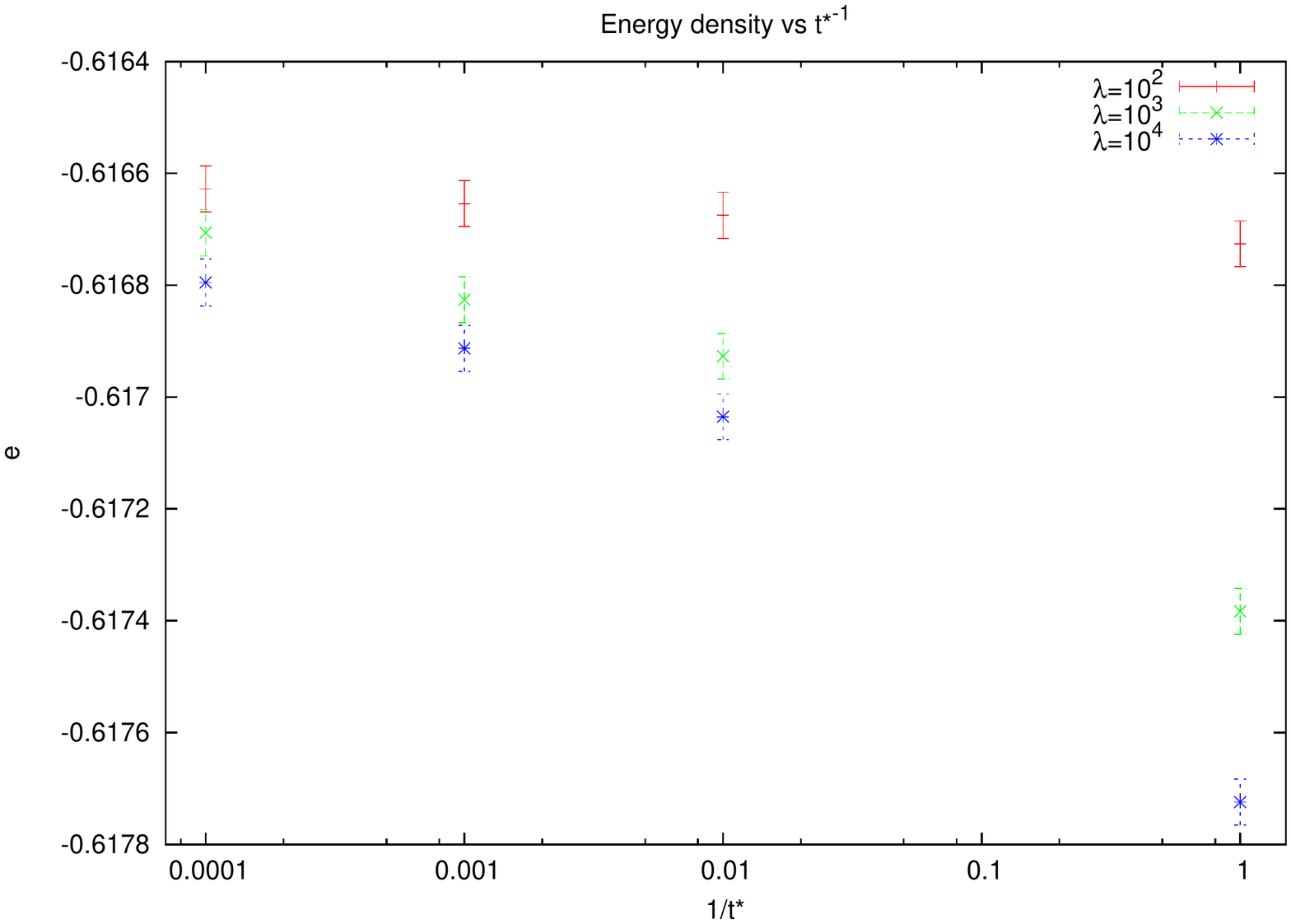}
\caption{The final energies one obtains with different $t^*$'s in a $L=100$ lattice. Each color indicates a different $\lambda$. Definitely the energy of the inherent 
structure is higher the greater $t^*$ is, no matter $\lambda$.}
\label{fig:lambda-tstar}
\end{figure}

We start by stressing the properties of $t^*$. Figure \ref{fig:tstarperde} shows how the
convergence time does not depend from this variable, but completely on $\lambda$.
We can also see (figure \ref{fig:lambda-tstar}) that, as intuition suggests, more 
Gauss-Seidel steps we perform before starting with $\lambda\neq0$ less we will be able 
to descend in energy: it is impossible for the Successive Over Relaxation to recover
the ``unwise'' greedy quench steps.

So if making a number of quench steps before does not decrease the convergence time, and
it entails an increase in the final energy, there is absolutely no point in keeping
$t^*\neq0$.

The situation is more delicate when we proceed to select an optimal $\lambda$. In this 
case the choice is not univocal, since it depends on which criteria we consider dominant.

\begin{figure}[!ht]
 \includegraphics[width=\textwidth]{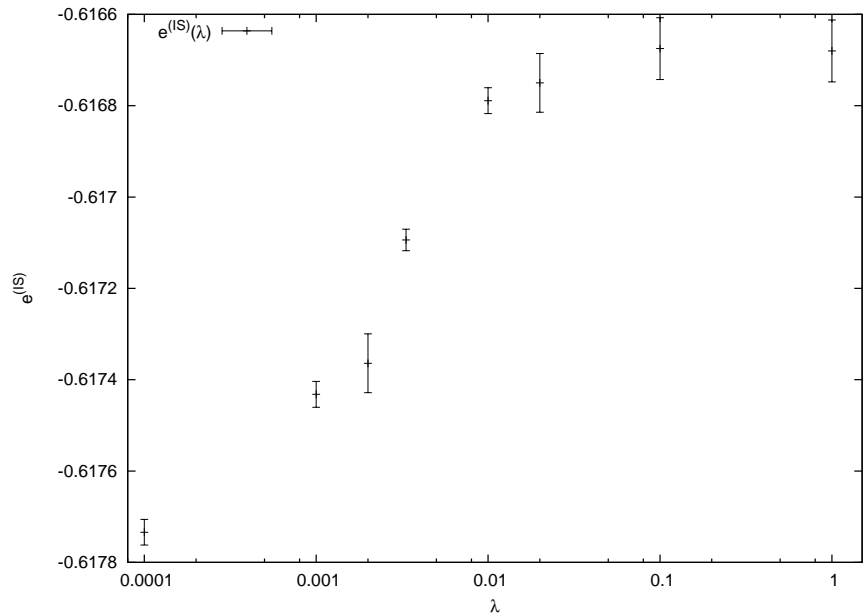}
\caption{The energy of the inherent structures in a $L=100$ lattice (except the point at $\lambda=300$ that comes from an $L=64$ lattice). 
We can ideally divide the plot in two sectors. In the left sector we have large $\lambda$'s: the dynamics takes place mainly at high energies 
and the system takes its time to choose the valley in which it wants to lay down. In the right sector, for smaller $\lambda$, the system dives 
quickly in a minimum, losing the choice of a lower valley. 
Mind that the dependency of $e^{(IS)}$ on the system size $L$ is studied in
Fig. \ref{fig:energy-L-lam}, below. With our statistical accuracy, the results seem
$L$-independent from $L=32$.}
\label{fig:scaling-lambda}
\end{figure}

Looking at the raw data one could be attracted by the possibility of a pretty scaling 
law for $\lambda$ such as, for example,
\begin{equation}
 e^{(IS)}(\lambda) = e^{(IS)} + A\lambda^{-A'}
,
\label{eq:scaling-lambda}
\end{equation}
from where it would be easy to extrapolate the energy of the inherent structure. 
Unfortunately it is not so direct to state the existence of a similar scaling law
(fig \ref{fig:scaling-lambda}), and also if it exists, it is not as simple as the 
one described in eq. \ref{eq:scaling-lambda}. Since the objective of this work is not
to find all the characteristics of our minimizing routine we did not linger in finding 
a scaling law.

\begin{figure}[!ht]
\begin{center}
 \includegraphics[width=\textwidth]{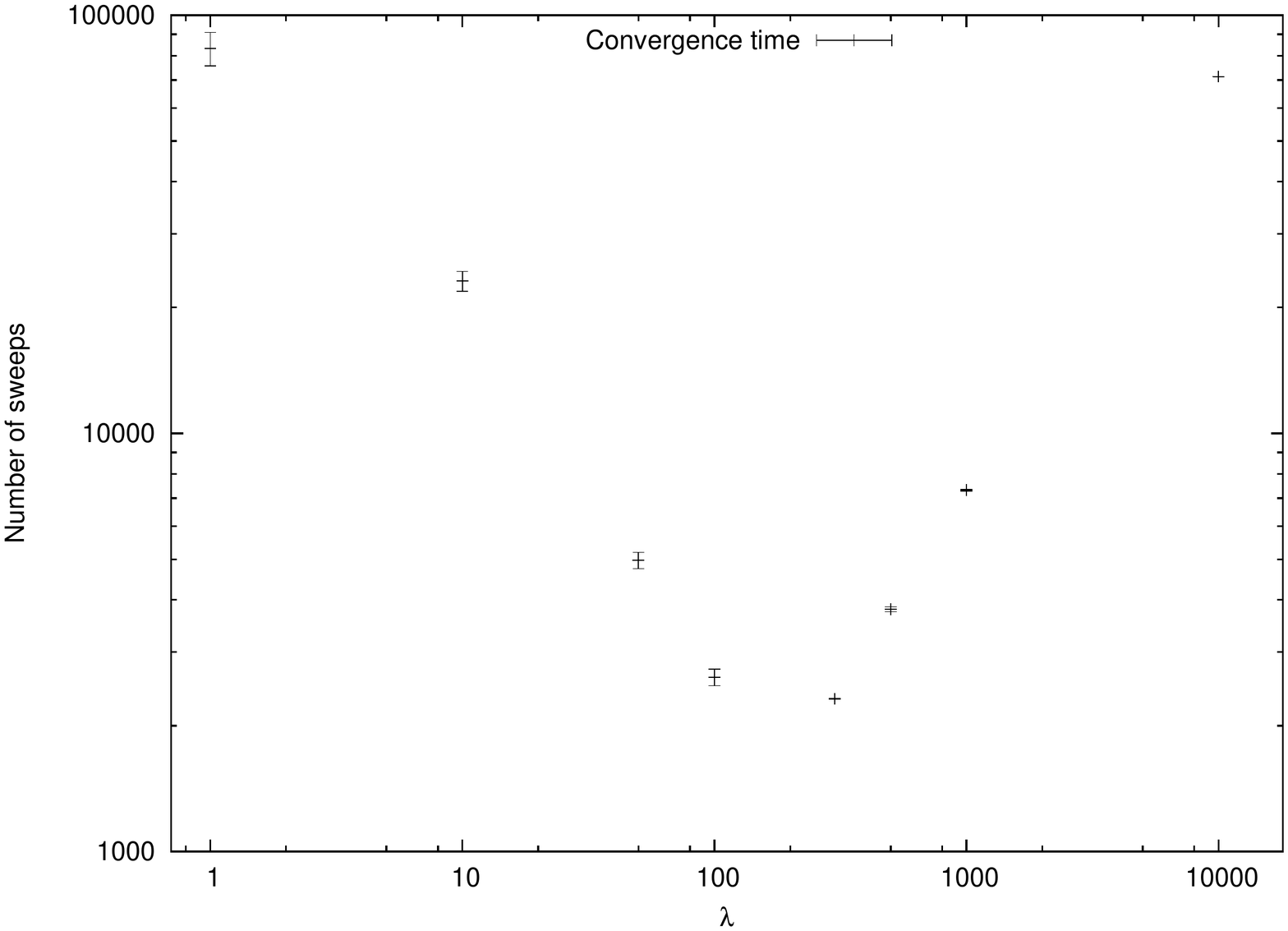}
 \caption{Convergence times for different $\lambda$. Here, as in figure \ref{fig:scaling-lambda}, it is possible to identify the two regimes. 
 For small $\lambda$ the system drops quickly into a minimum, but it takes it a long time to converge once in the valley. For large $\lambda$ 
 it stays plenty of time at high energies. We see that from the pure Gauss-Seidel the convergence time decreases linearly with $\lambda$, 
 because it becomes more and more effective once in a valley. On the other side, if we increase $\lambda$ too much the algorithm becomes too 
 ungreedy, loses too much  time at high energies, and the convergence time starts growing with $\lambda$.
 All the points come from $L=100$ systems, except $\lambda=300$ that is extracted from $L=64$.}
\label{fig:tconv}
\end{center}
\end{figure}

The other criterion to keep in mind when choosing our $\lambda$ is the convergence time,
which is absolutely not negligible. In fact although for small $\lambda$ the Successive
Over Relaxation makes the convergence faster, there is a $\lambda$ around $100$ (see fig. \ref{fig:tconv}) 
in which the algorithm becomes too ungreedy, and the convergence times start growing.
Indeed it is true that any $\lambda<10^5$ represents an objective gain respect to the
quench, since it yields lower energies in less time.

Nevertheless we ought to choose only one single $\lambda$ between the infinity available. Since
we did not believe that our algorithm could cross the phase transition for a specifically 
big $\lambda$\footnote{
If the algorithm were to cross the phase transition, the correlation lengths would
depend strictly on the system size for any $L$, since at the phase transition $\xi$ becomes infinite. On the
contrary, we see that the correlation functions collapse perfectly once the system is large enough (figures \ref{fig:CS-Tinf} and \ref{fig:CS_T019}), 
giving an $L$-independent correlation length $\xi$.}, 
and start finding sub-exponentially numerous low energy configurations, and
since the definition of inherent structure is intrinsically related to the minimizing
routine used, we had no problem choosing the $\lambda$ which made our simulations the 
fastest, for the rest of the investigation, except in the next section, where we wanted
to remark the differences (other than the energy) between configurations achieved with 
different $\lambda$'s or where we explicitly needed to achieve a very low energy.
Another good reason for having chosen $\lambda=100$ is that a greater one would have 
yielded inherent structures with too long correlation lengths, preventing us from being 
able to calculate inherent structures at the finite temperature we were interested in.
For those reasons we made our investigation with $\lambda=100$, comparing it with 
$\lambda=300$ to see if there is some physical difference between the regime in which 
the system passes most of the time at low energies and the one where it delays at
high energies.

\section{A comparison with genetic algorithms}
\label{sec:giappa}
A nice check for our new algorithm is comparison with different methods for finding low energy configurations.
The only work of this type that we found in literature is \cite{giappa}, which uses a hybrid genetic algorithm to 
find ground states in EA Heisenberg spin glasses with $\pm J$ couplings. In spite of those two differences, we thought
that it might be interesting to make a comparison with their results, so we made a little parenthesis in our work and
studied the $\pm J$ model, since the maximum lattice size analysed by them has been $L=13$ and this allowed us to
not spend too much CPU time in this intent.

\begin{figure}[!ht]
 \begin{center}
  \includegraphics[width=\textwidth]{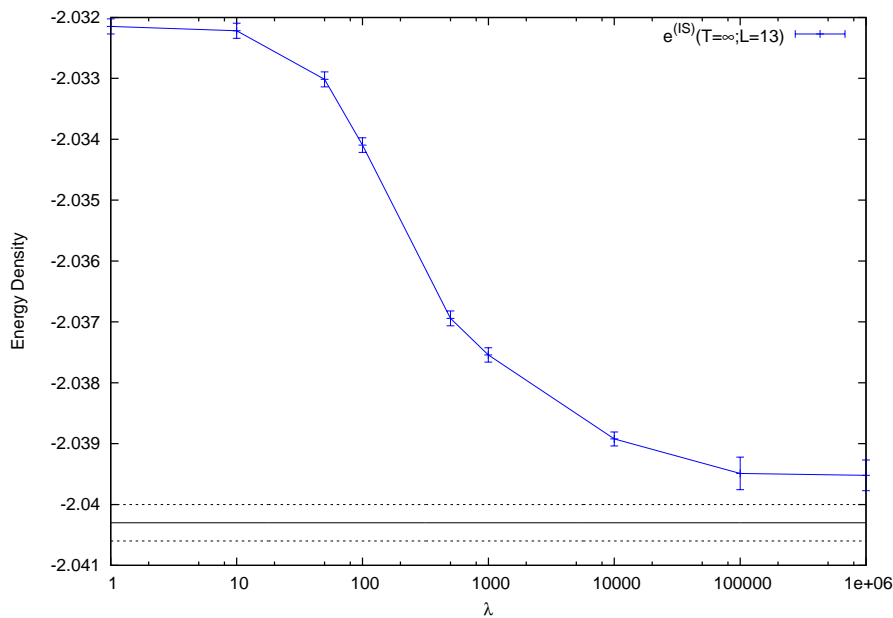}
  \caption{Inherent structures in the $\pm J$ Heisenberg model for $L=13$. The energy density is normalized in conformity with ref. 
  \ref{fig:japanL13}, i.e. without dividing by the number of dimensions. The blue points represent the average energies of the inherent structures 
  we found for each $\lambda$. The continuous black line represents $e_{GS}^{(GA)}$ obtained in ref. \ref{fig:japanL13} with a hybrid genetic algorithm, and the dotted lines its 
  uncertainty. 
  The difference between the energies achieved by their protocol and ours is fairly small, and they can't cope with system sizes greater than $L=13$,
  where finite-size effects are dramatic, so to our scope their algorithm would not be as useful as SOR, that converges quickly and permits us to 
  study very large lattice, keeping our measures in the thermodynamic limit.
}
  \label{fig:japanL13}
 \end{center}
\end{figure}

We made a series of simulations with growing $\lambda$ to see how close we could get to what they claimed to be
the ground state energy (we will call it $e_{GS}^{(GA)}$). As the reader can see in figure \ref{fig:japanL13} although we never 
reached $e_{GS}^{(GA)}$, the gap between our lowest energy and theirs is of the order of a standard deviation.

We also tried to reach lower temperatures by thermalizing the system at low temperatures and starting SOR from these configurations.
Although they were slightly lower, they still were not in the error bar of $e_{GS}^{(GA)}$:
\begin{enumerate}
 \item $e_{\pm J}(T=0.14, \lambda=10^5; L=13) = -2.0394 ± 0.0002$
 \item $e_{\pm J}(T=0.12, \lambda=10^5; L=13) = -2.0396 ± 0.0001$
 \item $e_{GS}^{(GA)}                         = -2.0403 ± 0.0003$
\end{enumerate}
It is to be noted that the correlation lengths of the inherent structures found after having thermalized at $T<\infty$ were appreciably 
different from those from $T=\infty$. This means that the situation is not so simple as it may appear, and a state cannot be characterized 
by the only energy.

We know that our relaxation routine is definitely not going under the phase transition, since the correlation lengths stay constant when we
increase the size of the lattice. 

If $e_{GS}^{(GA)}$ is really the ground state energy this means that the energy gap between the topological transition and the ground states
is very small and the topological transition does not have the importance one would like to give it. Nevertheless this would demonstrate that
the energy landscape is quite trivial until very close to the ground state, and a simple relaxation algorithm as SOR is enough to almost 
reach it.

It could also be, on the contrary, that $e_{GS}^{(GA)}$ does not indicate the ground state energy, but it is only a different estimate of the
energy at the topological transition. In this scenario the energy landscape would be quite non-trivial, proving that neither a hybrid genetic
algorithm can easily pass the topological transition.

Unfortunately in \cite{giappa} the only physical magnitude reported is the energy, so this does not
allow us to get to any conclusion. We can only say that SOR has revealed quite satisfying, since it yielded, with a competitive CPU effort,
energies at two standard deviations from what are claimed to be ground states.

\subsection{Inherent structures at infinite temperature}
The first physically relevant survey to do is the investigation of the inherent 
structures of the typical states of infinite temperature, doing SOR
from random configurations, since at $T=\infty$ all the configurations are equally 
probable.
\begin{figure}[!ht]
\begin{center}
\includegraphics[width=\textwidth]{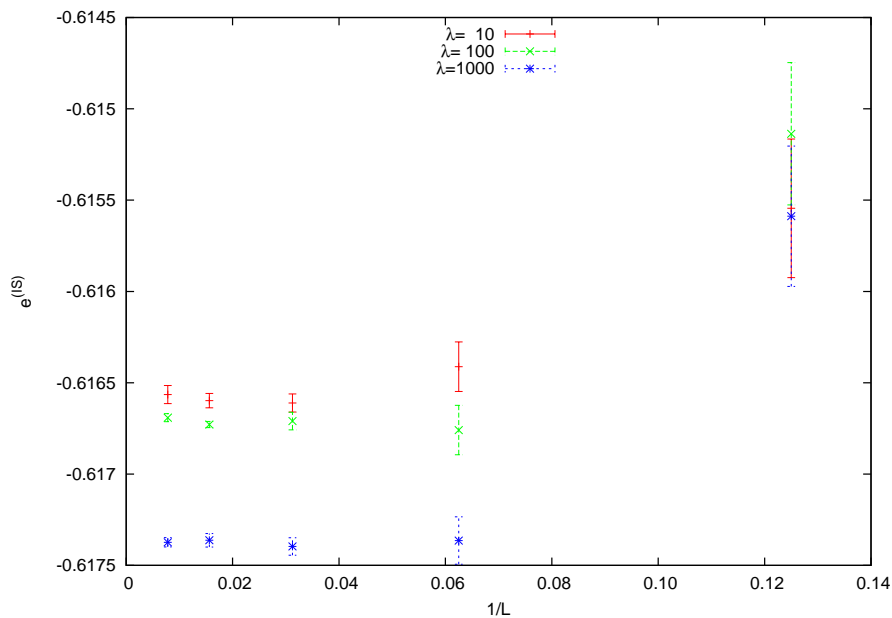}
\caption{The energy of the inherent structures for different system sizes. Already from $L=16$ this magnitude seems to become quite stable. Nevertheless
we're far from the thermodynamic limit, since we'll see that other quantities need bigger lattices.}
\label{fig:energy-L-lam}
\end{center}
\end{figure}
We stress immediately that, no matter the system size, for different 
$\lambda$'s not only the energy of the states is distinct (figure \ref{fig:energy-L-lam}), 
but also the correlation functions, consequently the correlation lengths and the magnetic 
susceptibilities, are different. This follows physical intuition, since a lower achieved energy corresponds to greater ordering. 
It is then clear how the concept of inherent structure is strictly binded to the routine we use to descend
in energy.

\begin{figure}[!ht]
  \begin{center}
  \includegraphics[width=\textwidth]{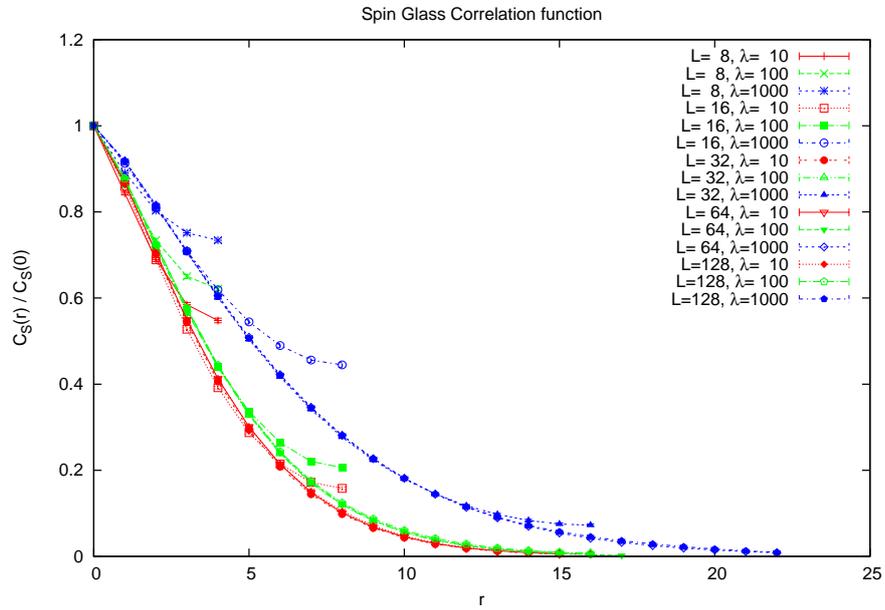}
  \caption{Normalized SG plane to plane correlation functions of the Inherent Structures for many $\lambda$'s and $L$'s at $T=\infty$. The data was truncated when it became smaller than three times its error,
  since at that point it was dominated by the statistic noise. 
  It appears clear from here that the configurations obtained depend strictly $\lambda$ since we can appreciate three different trends, one for 
  each $\lambda$. Moreover, this plot shows the strong finite-size effects for smaller lattices, since many correlation functions do not converge
  to zero. Notice how greater $\lambda$'s require larger lattices.}
  \label{fig:CS-Tinf}
  \end{center}
\end{figure}

Table \ref{table:xi} shows the typical values of the correlation lengths and magnetic susceptibilities one could extract from correlation 
functions such as the ones in figure \ref{fig:CS-Tinf}, for four different $\lambda$'s. As we just argued, it is not surprising that the correlation
lengths increase monotonically with $\lambda$.

\begin{table}[!ht] \caption{$\xi$ and $\chi$ from $T=\infty$, for different $\lambda$'s.} 
\centering 
\begin{tabular}{c|cccc} 
\hline\hline 
\\[0.3ex]
$\lambda$ & $\xi_{SG}$ & $\xi_{CG}$ & $\chi_{SG}$ & $\chi_{CG}$ \\ [0.5ex] 
\hline 
$10$      & $2.79\pm0.05$ &  $0.93\pm0.03$ & $178\pm3$     &  $0.0992\pm0.0011$\\ 
$100$     & $3.06\pm0.04$ &  $1.06\pm0.02$ & $205.3\pm1.4$ &  $0.1106\pm0.0010$\\
$1000$    & $4.49\pm0.07$ &  $1.57\pm0.05$ & $443\pm6$     &  $0.213\pm0.003$\\
$10000$   & $5.90\pm0.13$ &  $2.48\pm0.10$ & $927\pm22$    &  $0.432\pm0.021$ \\ [1ex]
\hline 
\vspace{0.5cm}
\end{tabular} 
\caption*{Table 4.1: Growth of the correlation lengths and the magnetic susceptibilities of the Inherent
Structures from $T=\infty$ for increasing $\lambda$. Higher $\lambda$'s yield lower energies which correspond to
more correlated configurations, as the data shows. All of the data comes from $L=128$ lattices, except the one for
$\lambda=10000$ that has $L=64$. Of course it suffers fair finite-size effects, since we see from figure \ref{fig:CS-Tinf} that
$L=64$ isn't enough also for $\lambda=1000$.}
\label{table:xi} 
\end{table}

Regarding the size of the system, from figure \ref{fig:energy-L-lam} we see that already
from $L=16$ the energy seems to be stable, so for $T=\infty$ we could consider ourselves
in the thermodynamic limit from $L=32$. Nevertheless we need to be careful, since energy 
is a quickly converging quantity. In fact from figure \ref{fig:CS-Tinf}
we see that for small $L$'s the correlation functions do not go to zero. 

This fact is 
confirmed by the binder cumulants, which only for larger sizes are compatible with their 
asymptotic values, as it is shown in table \ref{table:binder}, where we compare inherent
structures at infinite temperature for $\lambda=100$ for different system sizes.

\begin{table}[!ht] \caption{Binder Cumulants for $\lambda=100$ and $T=\infty$.} 
\centering 
\begin{tabular}{c|cc} 
\hline\hline 
\\[0.3ex]
$L$ & $B_L^{CG}$ &$B_L^{SG}$ \\ [0.5ex] 
\hline 
$8$      & $3.58\pm0.07$ &  $1.071\pm0.002$\\ 
$16$     & $3.34\pm0.09$ &  $1.179\pm0.004$\\
$32$     & $2.97\pm0.06$ &  $1.215\pm0.004$\\
$64$     & $3.05\pm0.06$ &  $1.232\pm0.004$\\
$128$    & $2.80\pm0.22$ &  $1.213\pm0.015$\\
$\infty$ & $3$           &  $1.\bar{2}$\\ [1ex]
\hline 
\vspace{0.5cm}
\end{tabular} 
\caption*{Table 4.2: The binder cumulants reflect the conclusions on the finite-size effects that we took when looking at the 
scaling of the energy with $L$. They tell us that from $L=32$ we are already in the thermodynamic limit, since they 
are compatible with their infinite limit. The uncertainty for 
$L=128$ is greater because we took less measures than with the other sizes, 
since there was no point in spending too much CPU time in data we did not really need for our investigation.
}
\label{table:binder} 
\end{table}

\newpage\null
\newpage

\section{Inherent structures at finite temperature}

To find inherent structures at finite temperature it was necessary to thermalize the 
system and only then go down in energy. Our ambition was to be always in the thermodynamic
limit, so that thermalization could become a very time-expensive task for low temperatures.
We chose temperatures from $T=0.19$ and greater because a previous article based on dynamical studies
\cite[M.Picco and F.Ritort (2004)]{picco} affirmed that at that temperature a phase transition was 
taking place. Also, we wanted our investigation to be in the thermodynamic limit, so
we could not descend too much in temperature because the closer to $T_{SG}$ we are, the
longer the correlation lengths are.

\begin{figure}[!ht]
\begin{center}
  \includegraphics[width=\textwidth]{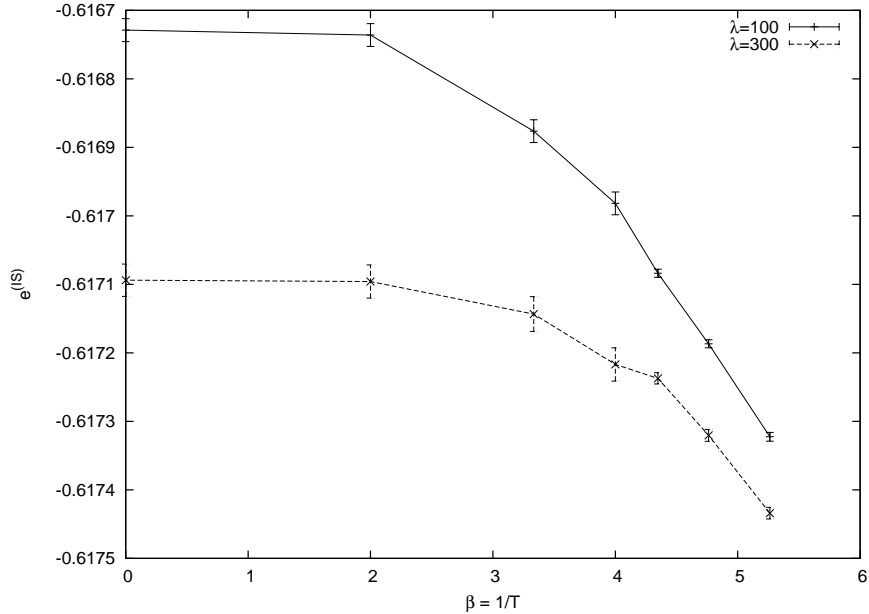}
  \caption{Energies of the inherent structures in function of the inverse temperature $\beta=1/T$ for two different $\lambda$'s. We plotted 
in function of $\beta$ to be able to show the infinite-temperature limit. The trend is qualitatively the same we encounter in supercooled 
liquids and structural glasses, where the energy of the inherent structure is almost independent from $T$ at high temperatures, and it 
descends steeply at lower temperatures.
Keep in mind that being in function of $\beta$ the energy descent at low temperature looks less steep than what it would with $T$ on the $x$-axis.
Nevertheless this lets us appreciate the extremely small variation between $T=\infty$ ($\beta=0$) and $T=0.50$ ($\beta=2$), since their error 
bars cross, in contrast with the appreciable descent for lower temperatures (higher $\beta$).
Notice how the difference between the two protocols decreases with decreasing temperature, so the choice of $\lambda$ becomes less relevant the
lower $T$ is.
The three points at lowest temperature (highest $\beta$) have less uncertainty than the others because the low-temperature data requested using 
$L=128$ while for the other we used $L=64$.}
  \label{fig:inh-str-beta}
\end{center}
\end{figure}

To get to know the thermalization time of the system we have waited until the energy became 
stationary and the three time-correlation functions described in section \ref{sec:corrtime} 
went to zero.

It turned out that although the thermalization time was not fast, it was not either 
prohibitive, so we have been able to thermalize $1000$ samples for each chosen 
temperature. For $T>0.23$ a $L=64$ lattice was enough, while for lower temperatures
the correlations grew too much and it was imperative to use $L=128$.

\paragraph{Trend of the Inherent Structure Energy}

Figure \ref{fig:inh-str-beta} shows how $e^{(IS)}$ varies with the inverse temperature $\beta=1/T$. It is almost constant
from $T=\infty$ to $T=0.5$, as the error bars of those two points overlap, and it starts descending
appreciably around $T=0.3$ ($\beta=3.\overline{3}$). This is the same qualitative scenario we encounter in
structural glasses \cite{sciortino}, where $e^{(IS)}$ is temperature independent for high $T$'s, and
it starts decreasing significantly only from some certain temperature.

\paragraph{Anharmonic Energy Term}

\begin{figure}[!ht]
\begin{center}
  \includegraphics[width=\textwidth]{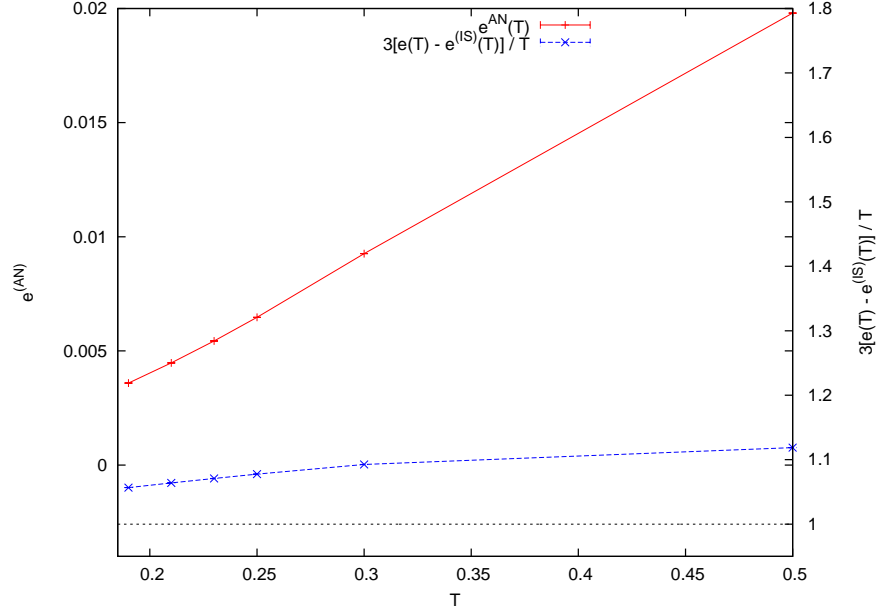}
  \caption{The anharmonic contribution to the energy. The red points refers to the left $y$-axis, and give the absolute value of the anharmonic
  term of the energy. The other lines refer to the  right $y$-axis. The blue points indicate the relative weight of the anharmonic contribution.
  The value $1$ indicates that it has no weight, and that is why there is a horizontal line indicating it. From the blue points we understand that 
  the anharmonic term is no larger than the $12\%$ at $T=0.5$, which is way over the critical temperature. At lower temperatures, though always in 
  the deep paramagnetic phase, the anharmonic contribution is less than half. This gives space to an interpretation of the energy of a state as
  principally an inherent structure plus harmonic thermal fluctuations.}
  \label{fig:e_anharmonic-multiplot}
\end{center}
\end{figure}
From the equipartition theorem we can divide the average energy at a given temperature in many contributions. There is a term given from the topology 
of the energy landscape, a harmonic term, proportional to $T$, that represents the thermal agitation around a minimum, and an anharmonic term that 
permits the movement through different valleys, which is more relevant the higher $T$ is. The relation is
\begin{equation}
 e(T) \simeq e^{(IS)}(T) + \frac{T}{3} + e^{(anh)}(T)
\label{eq:equipartizione}
\end{equation}
where $e^{(anh)}(T)$ is the anharmonic contribution to the energy and the factors $1/3$ derive from the energy normalization we chose.
We can compute the anharmonic contribution to the total energy by reversing equation \ref{eq:equipartizione}, and compute
\begin{equation}
 e^{(anh)}(T) = e(T) - e^{(IS)}(T) - \frac{T}{3}
\end{equation}
 that alone does not mean a lot to us, so we can understand its relative weight by looking at the quantity
\begin{equation}
 \frac{e(T)-e^{(IS)}(T)}{T/3} = 1 + \frac{e^{(anh)}(T)}{T/3}
.
\end{equation}

Figure \ref{fig:e_anharmonic-multiplot} shows the absolute anharmonic term of the energy on the $y1$-axis, and it shows its relative weight on
the $y2$-axis, where $y2=1$ marks the absence of an anharmonic contribution. We see that its weight does not exceed $12\%$ even when the temperature
is quite high, since $T=0.5\simeq5 T_{SG}$, hence it represents a non-primary effect. This fact lets us figure out that the energy of a configuration
is almost completely determined by the shape of the energy landscape plus harmonic thermal fluctuations over it. The small , but yet absolutely not negligible,
anharmonic effects are those that make us find different energies when we change the algorithm of energy descent, since they're the cause of its movement from one valley
to the other.

\begin{figure}[!ht]
\begin{center}
 \includegraphics[width=\textwidth]{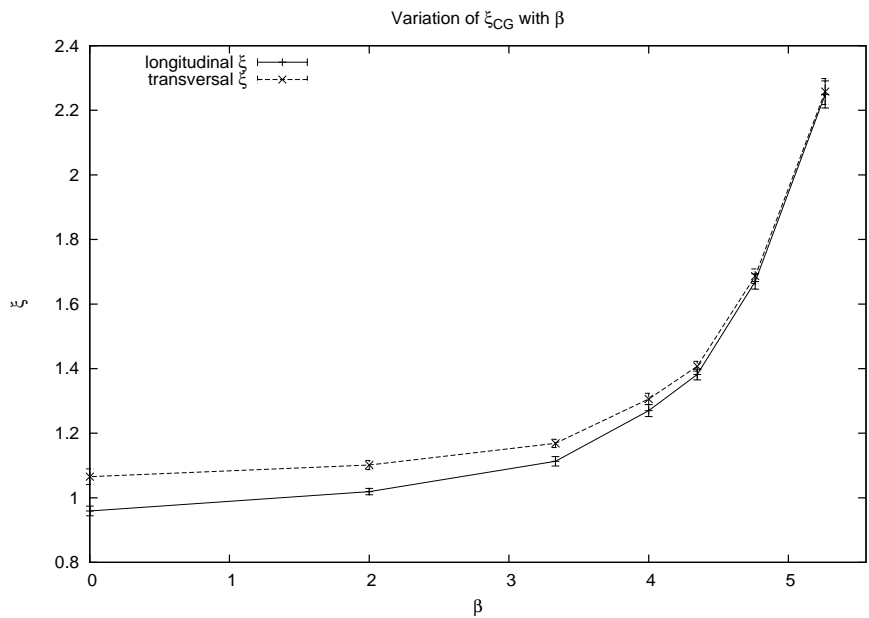}
 \caption{The longitudinal and transversal Chiral Glass correlation lengths of the Inherent Structures in function of the inverse temperature $\beta=1/T$. The trend is the 
 same for both, specially when going towards lower temperatures and the derivative starts growing consistently. 
 According to the most recent works $\beta_{SG} = 1/T_{SG} \simeq 7.75 $.}
\label{fig:xiCG_vs_T}
\end{center}
\end{figure}

\begin{figure}[!ht]
 \begin{center}
  \includegraphics[width=\textwidth]{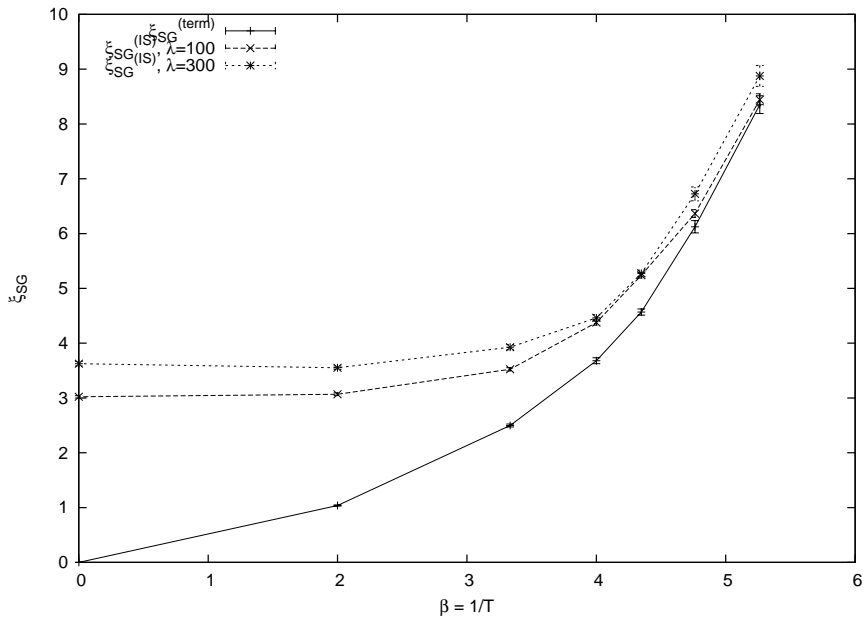}
  \caption{The  Spin Glass correlation length versus the inverse temperature $\beta=1/T$. There are both the correlation lengths of the 
  inherent structures reached with two different protocols, and of the thermal states. We see that although at infinite temperature 
  ($\beta=0$) they are consistently different, when $\beta$ grows they become more and more similar, up to the point that one could argue 
  that the divergence is due exclusively to the shape of the energy landscape. We see also how, lowering the temperature, the correlation lengths
  of the inherent structures depend less and less from the $\lambda$ used.}
  \label{fig:xi-IS-term_T}
 \end{center}
\end{figure}

\paragraph{Correlation lengths}
We also want to show (figures \ref{fig:xiCG_vs_T} and \ref{fig:xi-IS-term_T}) how correlation lengths vary with temperature.
From these trends we made extrapolations that we will show in a further section. By now we can appreciate the apparent divergence
that both present around $T\sim0.15$.

A more interesting fact that the correlation lengths can remark to us is that more we go down in temperature, more important the role of
the energy landscape becomes. In fact we can see in figure \ref{fig:xi-IS-term_T} that as we descend in energy the correlation length of a
thermalized configuration becomes more and more dominated by the one of its inherent structure. We can guess that the divergence of the
correlation length at the critical temperature is completely due to the underlying inherent structure scenario.

\begin{figure}[!ht]
 \begin{center}

  \includegraphics[width=\textwidth]{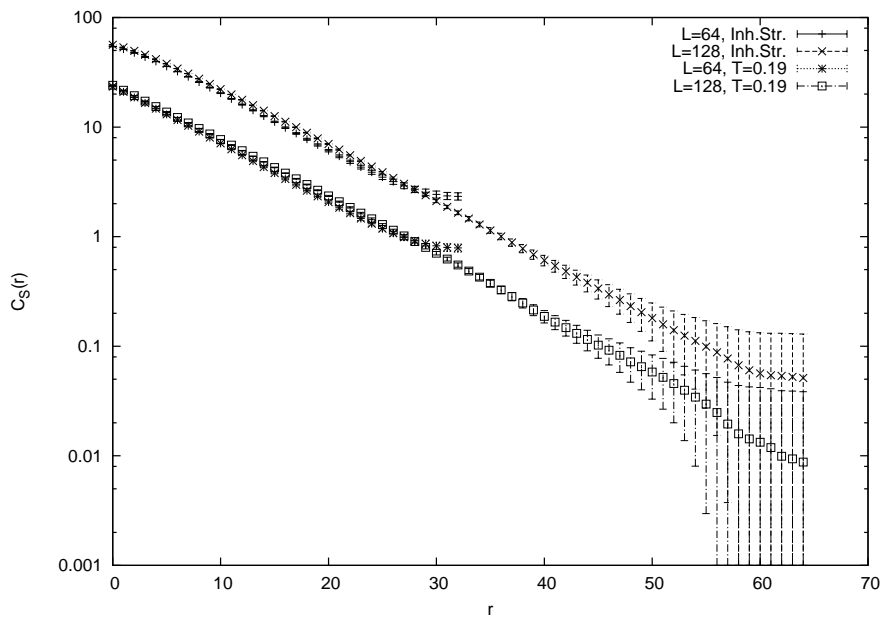}
  \caption{The correlation functions of the thermal state and its inherent structure at $T=0.19$. The similarity between the correlation lengths is
  quite evident, as also the exponential trend. We see that at $T=0.19$ the lattice size $L=64$ is not enough since the correlation function does not 
  go to zero (it is always far from being compatible with zero), while $L=128$ is enough because the last ten distances are compatible with zero.}
  \label{fig:CS_T019}

 \end{center}
\end{figure}

In figure \ref{fig:CS_T019} we show the spin glass correlation function for the states thermalized at $T=0.19$, and the same magnitude for their
inherent structures. The correlation lengths are indeed very similar since the difference between them is indiscernible by eye. We can also
appreciate the finite-size effects for the $L=64$ samples, since it is quite clear that the correlation lengths do not go to zero.


\begin{figure}[!ht]
 \begin{center}
   \includegraphics[width=\textwidth]{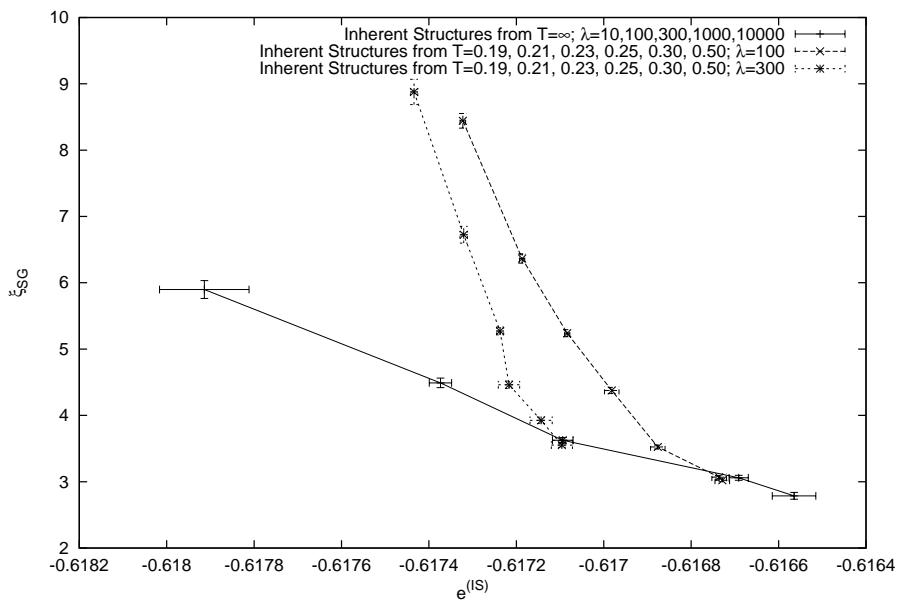}
  \caption{The correlation lengths of the inherent structures, associated to 
  their energy. The dashed lines represent inherent structures from $\lambda=100,300$ protocol, starting from states thermalized at different 
  temperatures. The solid line instead describes inherent structures from infinite temperature, with different $\lambda$'s. 
  If the energy were enough to characterize an Inherent Structure, each energy would have associated only one correlation length.
  On the contrary we see that same energies reached with different protocols have different $\xi$, so we can't sample all the inherent 
  structure population only performing SOR with different $\lambda$'s from one temperature, neither can we conforming ourselves to a single $\lambda$
  from different temperatures.
  All the data comes from $L=128$ lattices, except the points for $\lambda=100,300$ at $T=0.25,0.30,0.50$, which can be recognized because of their lower
  correlation length, and the point for $\lambda=10000$ and $T=\infty$, whose study was so slow that we had to limit ourselves to few samples,
  which were obtained from $L=64$ lattices.
  }
  \label{fig:notonlyE-xi}
 \end{center}
\end{figure}

\paragraph{Variety of the Inherent Structures and CG self-overlaps} 
By comparing the energies obtained with SOR from $T=\infty$ with a very high $\lambda$, with the ones obtained having first thermalized at finite
temperature, but with a lower $\lambda$, one could notice that it is possible to obtain the same energies in different ways. A naive thought could
be that these are simply two different ways to achieve the same result. This assertion would be incorrect because the energy is not the only 
relevant observable, and as we already said in section \ref{sec:giappa}, it does not characterize univocally a state. 
As a matter of fact magnetic susceptibilities, correlation lengths (fig. \ref{fig:notonlyE-xi}) and CG self-overlaps (fig. \ref{fig:selfqCG-IS})
may be very different for inherent structures wth the same energy, hence there is a whole variety of inherent structures for each energy.

\begin{figure}[!ht]
 \begin{center}
  \includegraphics[width=\textwidth]{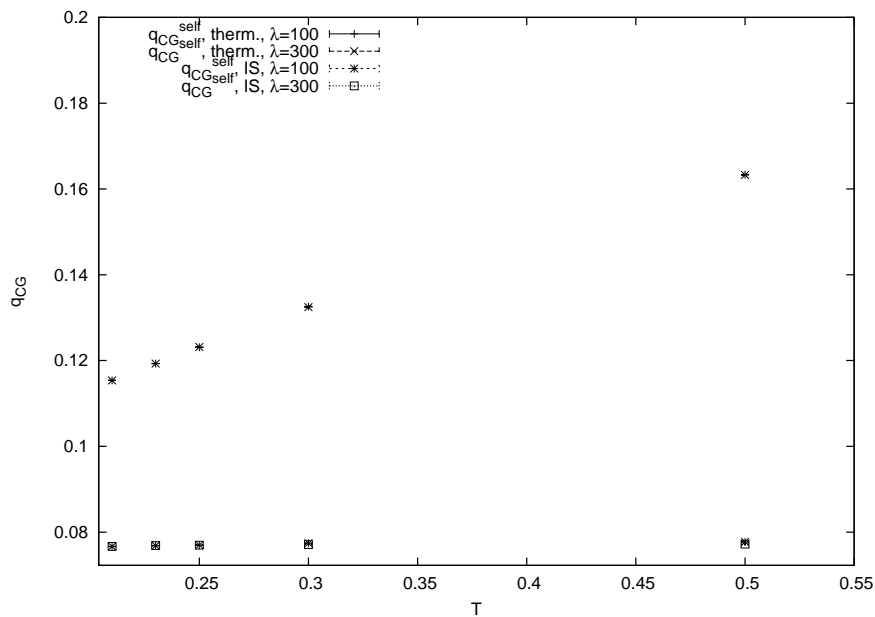}
  \caption{
  The Chiral Glass self-overlaps of the inherent structures compared with the ones of the thermal states. The Inherent Structures have 
  remarkable spin aligning properties, so, differently from Ising spin glasses, one could say whether a configuration can be a ground 
  state or not, by simple visual inspection.
  }
  \label{fig:selfovlpCG_T}
 \end{center}
\end{figure}

\begin{figure}[!ht]
 \begin{center}
  \includegraphics[width=\textwidth]{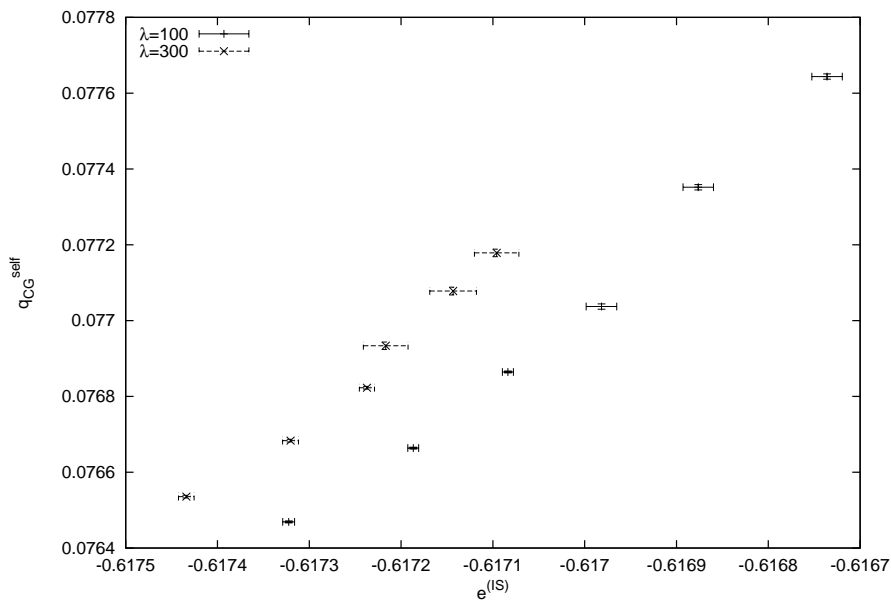}
  \caption{The CG self-overlap for Inherent Structures obtained with two different protocols. It appears clear that Inherent Structures with 
  the same energy have different self-overlaps, therefore the energy does not individuate univocally the Inherent Structures. The error bars 
  are greater in six of the points because they represent data coming from $L=64$ lattices, differently from the others that come from 
  $L=128$ (we had to use $L=128$ for lower $T$'s due to finite-size effects). Notice that lower $T$'s yield lower energies, so the same plot
  as a function of $T$ would look similar. For this reason the reader can notice that the self-overlap of the inherent structure decreases
  with $T$, and is not constant as it may appear from figure \ref{fig:selfovlpCG_T}}
  \label{fig:selfqCG-IS}
 \end{center}
\end{figure}

The CG self-overlaps deserve a separate discussion. It is commonly known that in Ising spin glasses there is no visual difference between a
minimum and a random configuration, so that visual inspection can not discern one from the other. On the contrary, we noticed that this is
not true in Heisenberg Spin Glasses. In fact, as one can see in figure \ref{fig:selfovlpCG_T}, the CG self-overlaps are very different between
thermal configurations and relative Inherent Structures. More specifically, we see that the CG self-overlaps are considerably smaller in the
Inherent Structures, which have greater alignment properties.

Although it might seem, from figure \ref{fig:selfovlpCG_T}, that the Inherent Structures' self-overlap is constant with temperature, one 
can correctly guess from figure \ref{fig:selfqCG-IS} that those too descend appreciably with $T$. Consequently we can characterize an
Inherent Structure also with its Chiral Glass self-overlap, and see, as it is well shown in that same figure, that this is another feature,
other than the energy and the correlation length, that distinguishes different kinds of Inherent Structure.

\section{Dynamics}
\begin{figure}[!ht]
 \begin{center}
  \includegraphics[width=\textwidth]{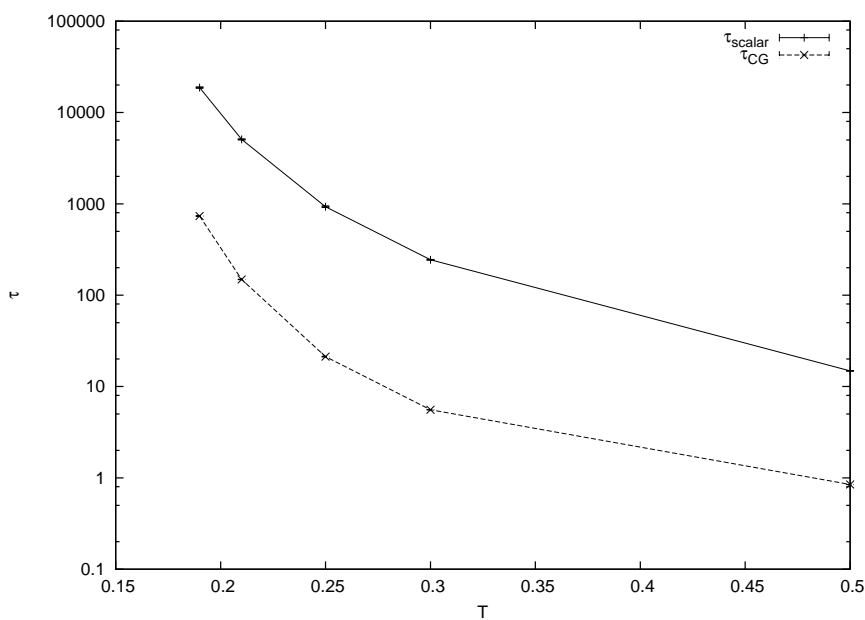}
  \caption{The two relaxation times $\tau_{Scalar}$ and $\tau_{CG}$. The chiral sector carries information for much less time.
  }
  \label{fig:tau_T}
 \end{center}
\end{figure}

The self-correlation times (figure \ref{fig:tau_T}) yield more or less the same result obtained in ref. \cite{picco}. They diverge at a temperature significantly greater than
$T_{SG}$ (or $T_{CG}$), justifying the analogy with structural glasses. Also in this case we postpone the interpolations to the next section. In any case the divergence seems quite predictable also
qualitatively, to a $T>T_{SG}\simeq0.12$. Notice how the Spin Glass signal is sensibly stronger than the Chiral Glass.

\section{Extrapolations}

Since most of the critical exponents of the 3-dimensional Heisenberg model are not known yet, we made some extrapolations that were possible with 
the data we collected. We have been able to estimate the critical exponents $\nu$ and $z$, the exponent of the divergence predicted by Mode 
Coupling Theory (MCT), and we also wanted to give some estimates of $T_{SG}$ by making extrapolations from far from $T_{SG}$, to show how
large lattices far from the critical temperature are effective as small lattices close to it, to make understand how it was possible to
underestimate it so much to think it zero, and specially to remark that the nature of the overestimation of $T_c$ with the $\tau$'s is not a
trivial misestimation effect due to being far from $T_{SG}$, since that effect would result in an underestimation.

\subsection{An Estimate of $T_{SG}$}
We tried to estimate, the spin glass critical temperature by means of the divergence of the Inherent Structures' correlation length. The scaling behavior near $T_{SG}$
should be
\begin{equation}
 \xi^{-1/\nu}\sim (T-T_{SG}),
\end{equation}

\begin{figure}[!ht]
 \begin{center}
  \includegraphics[width=\textwidth]{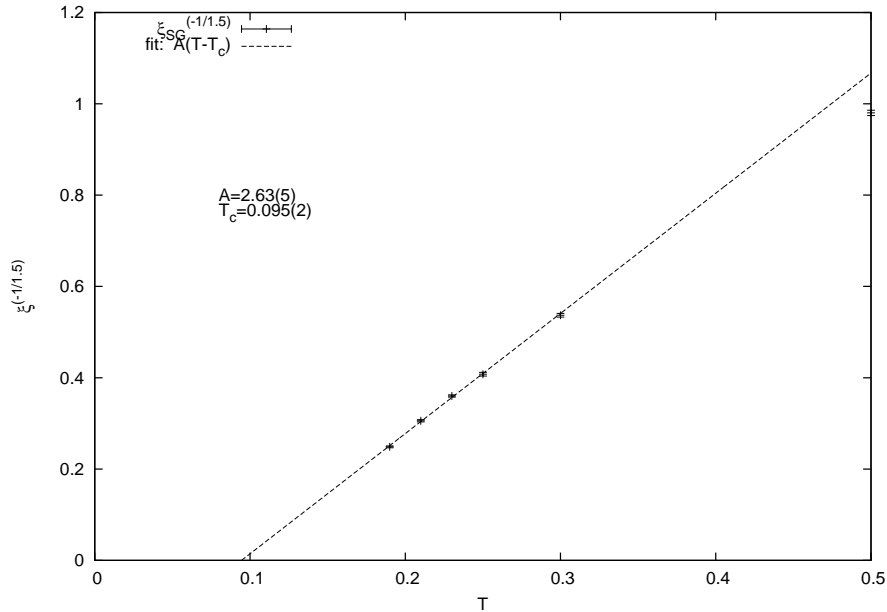}
  \caption{Power law fit of $\xi_{SG}^{-1/\nu}$ imposing the value estimated in \cite{tsgMF3} $\nu=1.5$. The point at highest temperature has been 
  excluded from the fit since it didn't respect the linear trend, and the scaling law is valid only close to $T_{SG}$. The intersect gives us $T_{SG}=0.95(2)$,
  that underestimates it considerably. We see how using very large systems is not effective if we look for critical quantities. It is quite better to work closer to $T_{SG}$ in spite of the finite effects, 
  since the scaling laws are valid only very close to $T_{SG}$.
  }
  \label{fig:xi_nu}
 \end{center}
\end{figure}
so if we plot $\xi^{-1/\nu}$ versus $(T-T_{SG})$, and impose $\nu$ to be the best estimate found until now, given in \cite{tsgMF3}, i.e. $\nu=1.5$, 
we can find $T_{SG}$ from the linear fit's intercept (figure \ref{fig:xi_nu}). Except for the point at $T=0.5$ the data fits quite well, but yet we 
find $T_{SG}^{(fit)}=0.095(2)$, $20\%$ lower than $T_{SG}=0.129$ reported in \cite{tsgMF3}. This result was expected, since we are estimating $T_{SG}$
with tools that are valid only near it, while our data are very far from it. Intuition leads to think that the data would assume curvature if we got
closer to $T_{SG}$, until reaching a stable slope, so probably the real value of $\nu$ greater than $1.5$.

Of course it is fairly reasonable that our estimate for $T_{SG}$ didn't reproduce the results obtained in works who focused on the phase transition.
This because the scaling laws are valid only close to $T_{SG}$, so for this aim it is worth to get some finite-size effects and work at smaller temperatures, 
around $T_{SG}$.

\newpage

\subsection{An Estimate of the Exponent $\nu$}
From the same data set of figure \ref{fig:xi_nu} we can try to deduce the exponent $\nu$, by fixing $T_{SG}=0.129$ as reported in \cite{tsgMF3},
and fitting on $\nu$. We plot $\xi^{-1}$ versus $T-T_{SG}$ and expect a trend of the form
\begin{equation}
 \xi^{-1}\propto ( T-T_{SG} )^\nu
,
\end{equation}

\begin{figure}[!ht]
 \begin{center}
  \includegraphics[width=\textwidth]{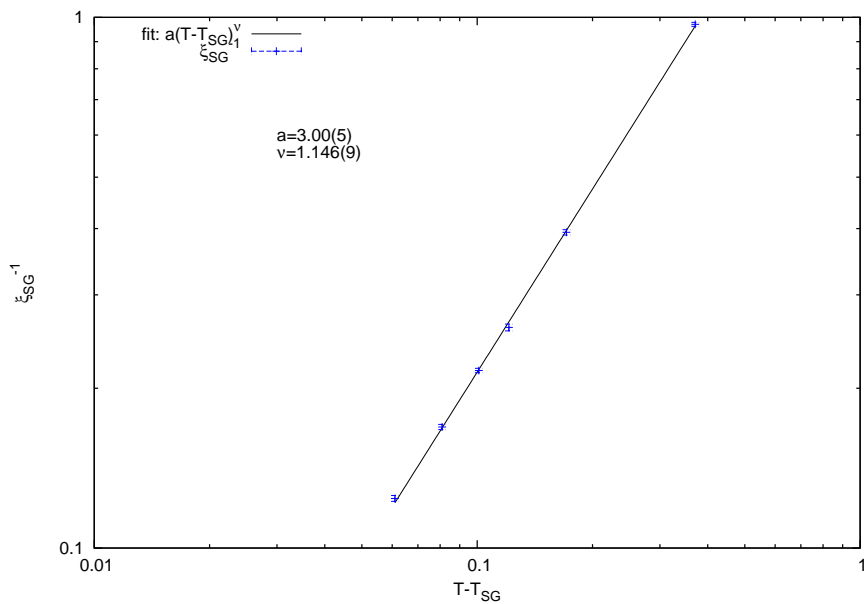}
  \caption{Imposing the $T_{SG}$ as the one evaluated in reference \cite{tsgMF3} ($T_{SG}=0.129$), we calculate the critical exponent $\nu$ with
  a fit. The result $\nu=1.146(9)$ is compatible with the one obtained in \cite{tsgMF3} with smaller lattices, but closer to $T_{SG}$. They obtain
  $\nu=1.01$ in a $L=8$, and $\nu=1.35$ with $L=12$.
  }
  \label{fig:xi_nu-findnu}
 \end{center}
\end{figure}
and obtain $\nu = 1.146 ± 0.009$ as it is shown in figure \ref{fig:xi_nu-findnu}. If we compare it with the results in \cite{tsgMF3} we see that they get $\nu(L=8) = 1.01 ± 0.02$ and
$\nu(L=12) = 1.35 ± 0.05$. Already $L=12$ with measures taken close to $T_{SG}$ gives a better estimate than ours. It appears clear how 
investigating the critical properties far from $T_{SG}$ is equivalent to going very close to the phase transition with very small lattices,
so are data is only good for an investigation of the pure paramagnetic phase.

\subsection{Self-correlation times}
This the most significant fit we present in this section, as it represents one
of the objectives of the present work. What we want to see is if the analogy with structural glasses holds and we can individuate a Mode
Coupling transition, which predicts a power law diverging trend of the relaxation times in the disordered phase, which is stopped, at lower 
temperatures, when we get in the region where activation leads the dynamics. The trend predicted by MCT is
\begin{equation}
 \tau \propto (T-T_c)^{-\zeta}
.
\end{equation}

\begin{figure}[!ht]
 \begin{center}
  \includegraphics[width=\textwidth]{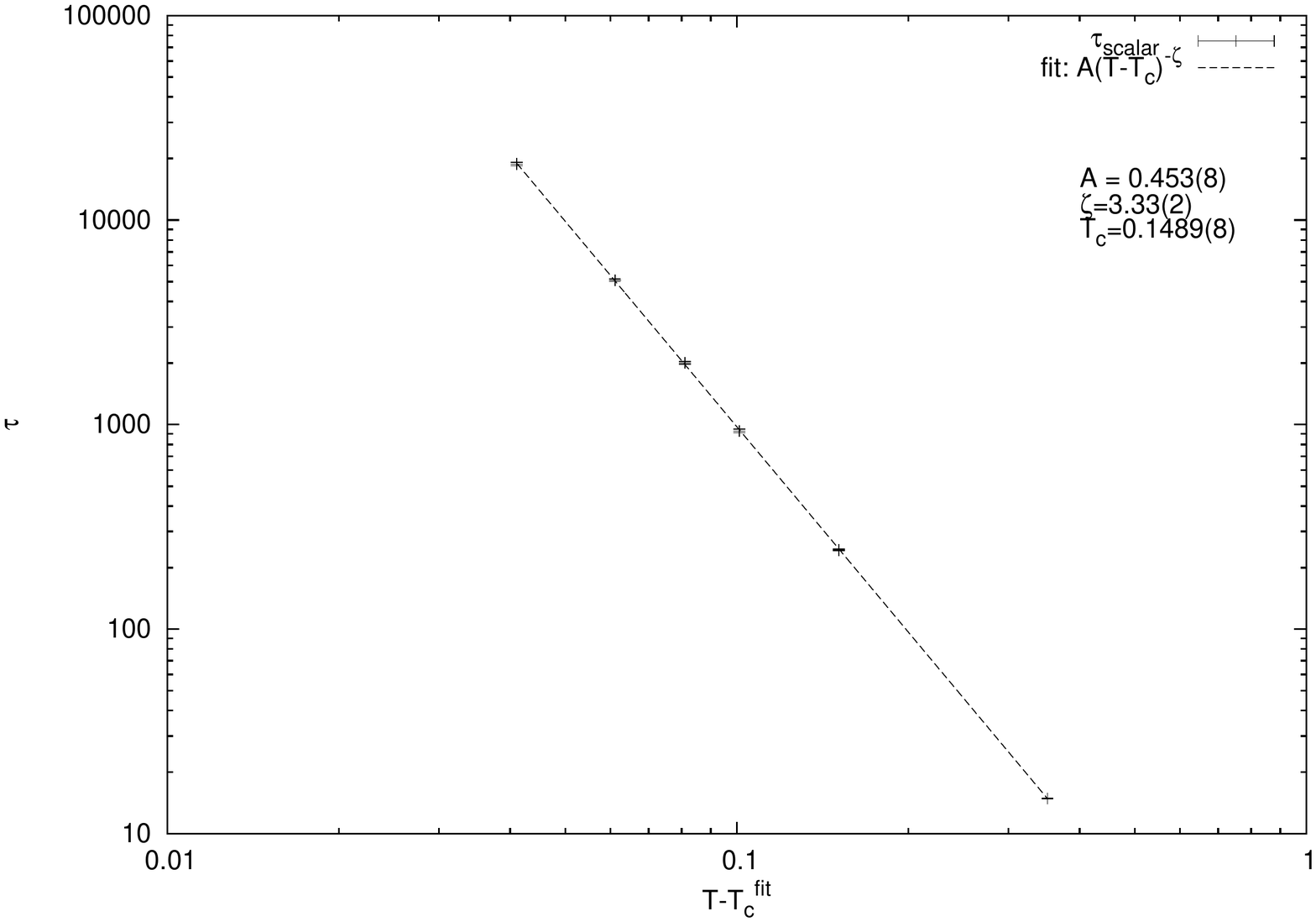}
  \caption{Power law fit of the scaling of the scalar relaxation times $\tau_{Scalar}$ in the paramagnetic phase. We encounter the same scaling, for $3$-$4$ orders of
  magnitude, in supercooled liquids. The scaling individuates the divergence of the $\tau_{Scalar}$'s at $T=0.1489(8)$, which far in the paramagnetic
  phase, just as predicted by Mode Couping Theory for supercooled liquids.
  }
\label{fig:tau_TTc}
 \end{center}
\end{figure}

In figure \ref{fig:tau_TTc} we can see how this prediction fits perfectly also for the EA Heisenberg model. We get a $T_c>T_{SG}$ that overestimates
the phase transition, just as MCT requests, and a perfect power law with exponent $B=3.33(2)$ which is of the same order of the those found by Kob
in \cite{kob} ($B=2.5 , 2.6$). We find $T_c=0.1489(8)$, that is lower than the one found in \cite{picco} ($T_c'=0.19(2)$), but only two standard 
deviations away. We stress that this $T_c$ shows itself as completely different from $T_{SG}$. In fact if it were different from
$T_{SG}$ just for the fact that we're very far from it, our result would at least be consistent with the one of the previous paragraph, underestimating
$T_{SG}$, while we find $T_c>>T_{SG}>>T_{SG}^{(fit)}$.

This result gives strong support to the possibility of a strong analogy between short-range interaction HSG and 
structural glasses.

\subsection{A Lower Bound for the Dynamic Exponent $z$}
Having data both for $\tau$ and for $\xi$ we can think of trying to estimate the exponent $z$, that relates them through
\begin{equation}
 \tau(\xi) = A\xi^z(1+B\xi^{-\omega}+...)
.
\end{equation}

\begin{figure}[!ht]
 \begin{center}
  \includegraphics[width=\textwidth]{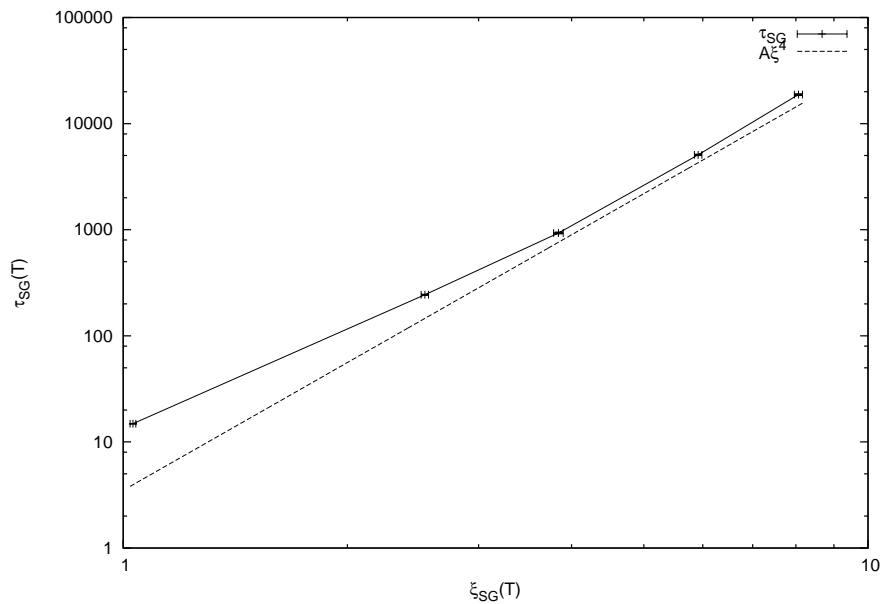}
  \caption{The relation between relaxation time $\tau_{Scalar}$ and correlation length $\xi_{Scalar}$ follows a power law with exponent $z$ close 
  to the critical point. Since our data was far from the critical point it was not linear enough (in the logarithmic scale) to permit a decent fit,
  we limited ourselves to giving a lower bound for this exponent. The points connected by a solid line represent our measures, and the dashed line 
  is the function $f(\xi)=A\xi^4$. It is clear how this function gives a loose lower bound of $z\geq 4$.
  }
  \label{fig:logg-tau-xi}
 \end{center}
\end{figure}

Unfortunately $\tau(\xi)$ has strong scale corrections, since we only have $\xi\leq8$, so we only feel to give a lower bound for $\xi$.
As we show in figure \ref{fig:logg-tau-xi} we can be reasonably sure that $z>4$. but we cannot say how much it will grow.

\newpage

\section{Minima or Saddles?}
To complete the frame that should indicate us if the analogy between the $3$-dimensional EA Heisenberg model and structural glasses is too hasty
we need to do a study of the density of negative directions in the hessian. In fact Mode Coupling Theory claims that the typical configurations
should be saddles until the temperature at which we have the topological transition. We need so to find if the typical configurations are closer to a
minimum or to a saddle point.

To this sake the algorithms we talked about until now are useless, since they are conceived to find minima of the potential energy and they
would not make us perceive the presence of saddles. We need to find a new routine that returns us the stationary configurations mindless of if they 
are stable (minima) or unstable (saddles).

The stationary configurations in our model are those for whom each spin $\vec s_{\vec x}$ is parallel to the local field $\vec h_{\vec x} = 
\sum_{y:||\vec x-\vec y||=1} J_{xy}\vec s_{\vec y}$. The problem of looking for both minima and saddle points then is equivalent to minimizing the
spins' component perpendicular to  $\vec h$, or maximizing the modulus of the parallel component. This way we could apply the algorithms we already 
used, to the energy functional
\begin{equation}
 H^{(saddle)} = -\displaystyle\sum_{\vec x} (\vec h_{\vec x}\cdot\vec s_{\vec x})^2
\end{equation}
and analyse the configurations out coming from this routine.

\newpage\thispagestyle{empty}\null\newpage 


\addcontentsline{toc}{chapter}{Conclusions}
\chapter*{Conclusions}
\markright{Conclusions}{}

We have made a systematic investigation of the behavior of the $3$-dimensional EA Heisenberg model in the paramagnetic phase in the thermodynamic limit, to see if we could 
remark features in common with structural glasses. Mode Coupling Theory predicts that the dynamics of structural glasses is dominated by the 
underlying structure of the energy landscape, so we made a deep study of the inherent structures of the model. 

For this sake the algorithms already 
used in literature were not enough effective, so we adapted the Successive Over Relaxation method to our problem and obtained a very performing 
routine. This algorithm depended on a parameter $\lambda$, whose modification influenced highly its convergence time and the properties of the
obtained inherent structures. We chose $\lambda$ who made our routine very fast, although it did not yield the lowest energies. This because we
did not believe that the properties of the configurations we found were intrinsically different from the others, and also because this choice permitted 
us to maintain ourselves in the thermodynamic limit since lower energies were related to longer correlation lengths. We noticed also that the energy is 
not enough to identify univocally an inherent structure, since same energies found with different routines lead to configurations with completely different
correlation lengths.

We analysed the dependence of 
the energy of the inherent structures from temperature, and remarked a trend that was qualitatively similar, with almost no dependence at high temperatures,
and a sudden decrease of their energies at lower temperature. Furthermore, we noticed that lowering the temperature the correlation lengths of
thermal configurations and inherent structures converge, so the divergence of the correlation length could be due exclusively to the energy 
landscape. An analysis of the anharmonic term of the energy showed that its contribution is fairly small up to pretty high temperatures, so it is not 
so crazy to schematize the energy of a thermal state as harmonic agitation over a local minimum.

We also noticed that the Inherent Structures of the Heisenberg Spin Glasses have particular alignment properties. In fact their Chiral Glass self-overlap
is consistently lower than the one of the thermal states, and decreases with the energy of the Inherent Structure. This feature would be very helpful to
recognize a Ground State, since it implies that, differently from Ising Spin Glasses, the minima of the energy are qualitatively different from random 
configurations.

Moreover, the data we had collected permitted us to make estimates for the critical exponents $\nu$ and $z$ of the model. Although they are not very reliable, 
since they come from states far from the critical temperature, where the scaling does not apply, they were useful to give some bounds that can be useful
for further reference, since in literature there still are very few attempts of evaluating the critical exponents of the EA Heisenberg model.
We focused on how working far from the phase transition with very large lattices is equivalent to using small systems close to the phase transition,
giving a measure of $T_{SG}$ that underestimated it consistently, just as it happened when this study was approached using too small lattices.

Later on we passed to the dynamic behavior of the model, since we wanted to see if we could observe the power law divergence of the relaxation time
in proximity of the dynamic transition, deep in the paramagnetic phase. Also this test had a positive feedback, since we observed the apparent divergence
of the relaxation time for 4 decades just as in structural glasses, the temperature at which the characteristic times would diverge was clearly greater 
than $T_{SG}$ and the scaling exponents of the same order of those obtained for supercooled liquids.

All this data is consistent with the identification of the EA Heisenberg model as a possible lattice model with the same universality class of
structural glasses. If this were true it would be very useful to study in an alternative way structural glasses.
Further studies to make more solid this impression, would refer to the investigation of the configurational entropy of the model, and, work that 
is already in progress, a detailed analysis of the density of negative directions in the hessian matrix, to see if we can individuate the topological
transition as its vanishing, with the arising of an activated dynamics. If those tests too will give a positive result it will be then enough evidence
to start considering a rigorous, more engaged, theory of our conjectures.

\newpage\thispagestyle{empty}\null\newpage 
\appendix
\chapter{Appendix}

\section{Fourier Correlation length}
\label{app:fourier}
We show how to get to the formula \ref{eq:xi-fourier} for the correlation length, and show a computationally efficient way to
calculate its terms.

Given a generic field $\phi(\vec x)$\footnote{For the purpose of our work $\phi$ can represent both th Chiral Glass overlap $e^\mu$
and the Spin Glass overlap $\tau_{\alpha\beta}$.} the action for a free Hamiltonian is
\begin{equation}
 S=\frac{1}{2}\displaystyle\int d^3x\{(\vec\nabla\phi(x))^2+\bar{m}^2\phi(x)^2\},
\end{equation}
from which descends the propagator
\begin{equation}
 \hat{G}(\vec k) =  \frac{Z(\bar{m}^2)}{k^2+\bar{m}^2}\hspace{1cm},\hspace{1cm}k^2=4\displaystyle\sum_{\mu=1}^D\sin^2(\frac{k_\mu}{2})
\end{equation}
where $Z(\bar{m}^2)$ is an unknown function of the mass $\bar{m}$. We know that the propagator $\hat G$ is the Fourier Transform of the correlation
function 
\begin{equation}
\label{eq:gr}
G(\vec r) = \frac{1}{N}\displaystyle\sum_{\vec x}<\phi(\vec x)\phi(\vec x +\vec r)> 
.
\end{equation}
We can obtain the correlation length $\xi^2=1/m^2$ by making the ratio between the $\hat G$ calculated for two different 
impulses $k$. This way we would get rid of the factor $Z(m^2)$. To this sake we use $\vec k=0$ and $\vec k_{min}=(2\pi/L,0,0)$, 
that is the minimum impulse with periodic boundary conditions, and we consider the ratio between
\begin{equation}
 F    = \hat G (\vec k_{min}) = \frac{Z(\bar{m}^2)}{4\sin^2(\frac{\pi}{L})+\bar{m}^2}
\end{equation}
and the magnetic susceptibility
\begin{equation}
 \chi = \hat G (      0     ) = \frac{Z(\bar{m}^2)}{\bar{m}^2}
,
\end{equation}
that yields directly
\begin{equation}
\label{eq:app-xi-f}
  \xi_F^2 = \frac{1}{4 sin^2 (\frac{\pi}{L})} [\frac{\chi}{F} - 1]
.
\end{equation}
Just to stay connected with the literature, we stress that with a finite differences approximation the correlation length in eq. \ref{eq:app-xi-f},
can be written in differential form, since
\begin{equation}
 \xi_F^2 = \frac{1}{\vec k_{min}^2} [\frac{\hat{G}(\vec 0) - \hat{G}(\vec k_{min})}{\hat{G}(\vec k_{min})}] 
       = -\frac{1}{\hat G(\vec k_{min})}\frac{d G}{d\vec k^2}\arrowvert_{\vec k=0}
\end{equation}

\paragraph{Calculating $F$ and $\chi$}
An operative way of calculating  $F$ and $\chi$ is made very simple thanks to the convolution theorem. Infact we see from equation \ref{eq:gr}
that the correlation function is a convolution product of the field $\phi$ with itself, hence its Fourier Transform $\hat G$ is equal to
\begin{equation}
 \hat G (\vec k) = \frac{1}{N}|\hat\phi(\vec k)|^2
,
\end{equation}
where $\hat\phi(\vec k)$ is the Fourier Transform of the field $\phi(\vec x)$. Remembering that we're only interested in its value at 
$\vec k=0,\vec k_{min}$, we can use an explicitly computable form for $\hat\phi(\vec k)$ for $F$ and $\chi$, helped by the fact that
$\vec k_{min}$ is parallel to the $x$-axis
\begin{equation}
 \hat\phi(\vec k_{min}) = \displaystyle\sum_{n_x}e^{i2\pi n_x/L}\sum_{n_y,n_z}\phi( (n_x,n_y,n_z) ) 
			= 	       \sum_{n_x}e^{i2\pi n_x/L} P(\vec x) 
\end{equation}
where $n_x,n_y$ and $n_z$ are the components of the position vector $\vec x$, and we recognize the plane variables $P(\vec x)$ defined in
section \ref{subsec:corr}.

We now have a numerically simple way to express $F$, from which the one for $\chi$ descends trivially,
\begin{eqnarray}
\chi &=& \frac{1}{N}\displaystyle(\sum_{\vec x}\phi(\vec x))^2\\
F    &=& \frac{1}{N}\displaystyle( \displaystyle\sum_{n_x}e^{i2\pi n_x/L}\sum_{n_y,n_z}\phi( (n_x,n_y,n_z) ) )^2
.
\end{eqnarray}

\section{Integral Estimators}
\label{app:estimators}
We want to show that in the thermodynamic limit the integral estimators $\xi^{(02)}$ are exactly equal to the correlation length $\xi_F$.
We remind its definition
\begin{equation}
 \xi^{(02)}=\sqrt{\frac{J_2}{\chi}}\hspace{1cm},\hspace{1cm}J_2=\displaystyle\sum_{r=1}^{L/2}r^2 C(r)
\end{equation}

\begin{figure}[!th]
 \begin{center}
  \includegraphics[width=\textwidth]{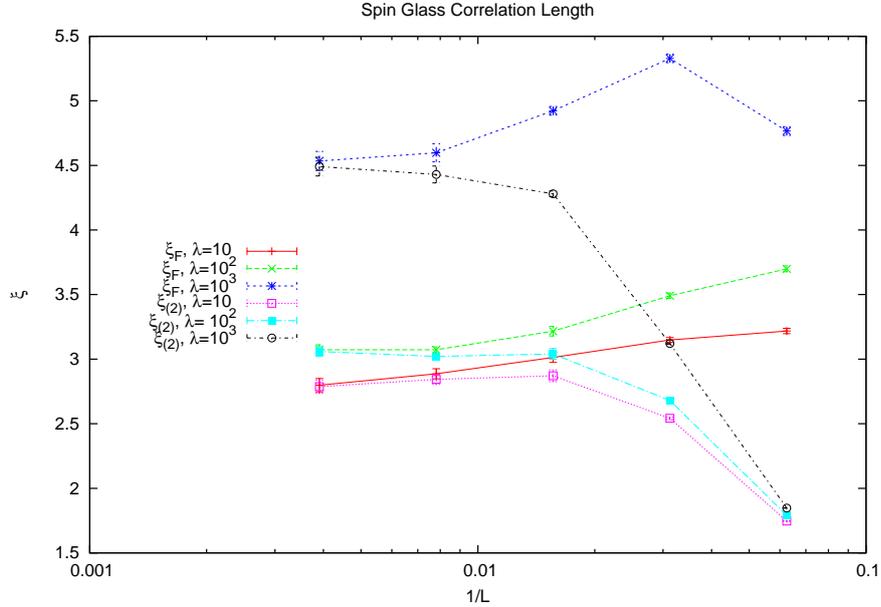}
\caption{Correlation lengths of configurations obtained with $3$ different protocols ($\lambda=10,100,1000$). We see how in small lattices,
far from the thermodynamic limit, $\xi_F$ and $\xi^{(02)}$ belonging to the same type of configuration are very different, since their 
equality is valid only in the thermodynamic limit. With growing system sizes the two correlation lengths start converging considerably.
The convergence of these two correlation functions can be used as index to establish if the thermodynamic limit is valid or not.}
\label{fig:xi02-xiF}
 \end{center}
\end{figure}

The minimum impulse propagator, keeping in mind that we work with periodic boundary conditions, can be rewritten as
\begin{equation}
 F = C(0) + C(\frac{L}{2}) + \displaystyle\sum_{r=1}^{L/2-1} C(r)\cos(\frac{2\pi r}{L})
\end{equation}
where the sum on the sines disappears because we're summing an odd function over a symmetric interval.
If we pass to the large $L$ limit, in which we want to demonstrate the identity, we can develop the cosine for small angles
and neglect the term $C(\frac{L}{2})$, since $C(r)$ vanishes at long distance. Hence
\begin{eqnarray}
 F &=& C(0) + 2\displaystyle\sum_{r=1}^{L/2-1} C(r) [1 - \frac{2\pi^2r^2}{L^2} + o(\frac{r^4}{L^4})] =\\
   &=& \chi - (\frac{2\pi}{L})^2 J_2 + o(\frac{J_4}{L^4})
.
\end{eqnarray}
We can insert this expression in the formula for $\xi_F$ getting
\begin{equation}
\xi_F^2 = \frac{1}{4 sin^2 (\frac{\pi}{L})} [\frac{1}{1-(\frac{2\pi}{L})^2 \frac{J_2}{\chi}+o(\frac{J_4}{L^4 \chi})} - 1] 
,
\end{equation}
from which if we develop another time for large $L$, keeping in mind that for small $x$ we have $\frac{1}{1-x}\simeq1+x$, 
and take the limit $L\rightarrow\infty$ we obtain finally
\begin{equation}
 \displaystyle\lim_{L\rightarrow\infty} \xi_F = \sqrt{\frac{J_2}{\chi}} \equiv \xi^{(02)}
.
\end{equation}
Figure \ref{fig:xi02-xiF} shows how the correlation lengths $\xi_F$ and $\xi^{(02)}$, that differ quite much for small system size, 
converge as the system approaches the thermodynamic limit.

\section{Asymptotic values of Binder's Cumulants}
\label{app:binder}
We show how to extract the asymptotic value $B_\infty^{SG}$ of the SG Binder Cumulant $B_L^{SG}$ defined in section \ref{subsec:binder}
in the case of configurations at infinite temperature (random configurations).
The calculation of $B_\infty^{CG}$ is analogous so it won't be shown. We remind the definition
\begin{equation*}
 B_L^{SG} = \frac{<tr(QQ^\dagger)>^2}{<tr(QQ^\dagger)^2>} = \frac{<q^2>^2}{<q^4>}
\end{equation*}
and proceed calculating separatly numerator and denominator for spins with $m$ components.
We start with the numerator
\begin{eqnarray*}
<tr(QQ^\dagger)> &=& <\displaystyle\sum_{\alpha,\beta}^m Q_{\alpha\beta}Q_{\alpha\beta}^\dagger>=\\
 &=&\sum_{\alpha,\beta}^m \sum_{x,y}<s_{x,\alpha}^{(a)}s_{x,\beta}^{(b)}s_{y,\alpha}^{(a)}s_{y,\beta}^{(b)}> =\\
 &=& \sum_{x,y}<(\vec s_x^{(a)}\cdot\vec s_y^{(a)})><(\vec s_x^{(b)}\cdot\vec s_y^{(b)})> = \\
 &=& \sum_{x,y} \delta_{xy} = \\
 &=& N
\end{eqnarray*}
and proceed with the denominator, where due to the square we have two more position indexes
\begin{eqnarray*}
 <tr(QQ^\dagger)^2> &=& \displaystyle\sum_{x,y,z,t}<(\vec s_x^{(a)}\cdot\vec s_y^{(a)})(\vec s_x^{(b)}\cdot\vec s_y^{(b)})
						 (\vec s_z^{(a)}\cdot\vec s_t^{(a)})(\vec s_z^{(b)}\cdot\vec s_t^{(b)})> =\\
 &=&\sum_x <\vec s_x^{(a) 4} \vec s_x^{(b) 4}> + \\
 &+&\sum_{x,z}<\vec s_x^{(a) 2}\vec s_x^{(b) 2} \vec s_z^{(a) 2}\vec s_z^{(b) 2}> + \\
 &+&2\sum_{x,y}<(\vec s_x^{(a)}\cdot\vec s_y^{(a)})^2><(\vec s_x^{(b)}\cdot\vec s_y^{(b)})^2>
.
\end{eqnarray*}
We compute the average $<(\vec s_x\cdot\vec s_y)^2>$ by putting ourselves in the reference frame of $\vec s_x$, aligning to it our first
axis. This way the scalar product is equal to the first component of $\vec s_y$. Hence
\begin{equation*}
 <(\vec s_x\cdot\vec s_y)^2> = \vec s_{y,1}^2 = \frac{1}{m}
\end{equation*}
since $\sum_{\alpha=1}^m \vec s_{y,1}^2 =1$ and there is no preferential direction. Therefore, summing up all terms we have
\begin{equation}
 <tr(QQ^\dagger)^2> = N + N(N-1) + \frac{2}{m^2}N(N-1)
.
\end{equation}
Recomposing the cumulant and performing the thermodynamic limit we get
\begin{equation}
 B_L^{SG} \stackrel{N\rightarrow\infty}{\longrightarrow} 1 + \frac{2}{m^2}
\end{equation}
So in our case $m=3$ we have $B_\infty^{SG}=\frac{11}{9}=1.\bar{2}$.

\backmatter

\addcontentsline{toc}{chapter}{References} 
\bibliographystyle{mprsty} 
\bibliography{tesi}

\newpage
\thispagestyle{empty}
\null
\newpage


\addcontentsline{toc}{chapter}{Acknowledgements}
\chapter*{Acknowledgements}
\markright{Acknowledgements}{}

Ci sono molte persone che vorrei ringraziare,
per una volta che posso farlo in maniera ufficiale.
In fondo la maggior parte della gente legge solo questi, 
quindi tanto vale essere prolissi.

\verse{
\textit{
V\'ictor, 
valido, serio e disponibile,
gli sono riconoscente per avermi accolto nel suo gruppo ed
aver avuto la pazienza di dedicarsi a me,
insegnandomi moltissimo nella mia esperienza a Madrid.
Mai avrei sperato di trovare qualcuno migliore.
Insieme a lui tutto il gruppo, che ha reso la mia permanenza
estremamente piacevole e stimolante.\\
Bea, che si è adoperata da subito per farmi trovare bene anche al di
fuori dell'università.\\
David e Luis Antonio, estremamente capaci, sempre pronti a dare una mano
in ogni evenienza.
}

\textit{
Federico, mi ha 
dato la possibilità di fare questa esperienza, seguendomi
a distanza come anche aveva fatto per la triennale.
}

\textit{
Aurora,
mi è stata a fianco in ogni momento,
e se non dimenticherò quest'anno
è merito suo e delle sere passate
con John Boy, cachimba, cioccolate calde,
film e plan A, oltre che del suo essenziale supporto
nei momenti di difficoltà.
}

\textit{
Samu,
anche se in due non facciamo un neurone,
non mi succedeva da tempo di trovarmi
così in sintonia con qualcuno 
in così poco tempo.
Dev'essere la mancanza di cervello.
}

\textit{
Davide,
questo percorso l'ho fatto insieme a lui,
di pari passo in ogni sua fase.
Dal laboratorio di fisichetta, a scommettere sul
numero di conteggi dei contatori,
alla pressione per la consegna della tesi di 
laurea in ritardo, come al solito.
}

\textit{Giulio e Valerio,
quasi non si conoscono di persona,
ma sono accomunati dal fatto di rappresentare per me l'esempio da seguire, 
lo stereotipo del fisico autentico, senza chiodi nel cervello.
}

\textit{
Sergino e Pablo,
si sono adoperati perchè il mio percorso formativo non fosse frutto del caso,
per farmi affrontare le mie scelte con consapevolezza,
facendomi ora avere le idee chiare su come intavolare il mio futuro,
ed insegnandomi che un cavallo vincente non si gioca mai piazzato.
}

\textit{
La mia famiglia più stretta, in quanto famiglia,
quella con cui ho condiviso quasi tutte le migliori esperienze,
che mi conosce meglio di quanto possa conoscermi io stesso, 
senza la quale non amerei la mia vita come adesso, 
non sarei lo stesso e non sarei migliore,
che mi ha donato esperienze che invidio al mio passato,
mi ha sempre sostenuto o perdonato quando ne avevo bisogno,
mantenendo sempre una stima smisurata, spesso ingiustificata, di me.
\\Fausto e Anna, mi avete sempre dato la sicurezza di sapere a chi rivolgermi nel momento del bisogno, in qualsiasi momento.
\\Mauro e Silvia, della mia infanzia ho più ricordi in casa vostra, che rimane tutt'ora un punto di riferimento fisso, che altrove.
\\Sergino e Giorgia, il futuro passa per casa vostra, e ormai anche le vacanze in montagna.
\\Stefano, la competizione per il ruolo \emph{alpha} rimane aperta, anche se per ora ti lascio un po' di vantaggio.
\\Valeria, ricordo di quando ti dissi che la mia seconda madre è la Luna, ma la terza sei tu.
\\Azzurra, come si fa-ha-ha la doccia pa-ha-zza? E come si diventa cugini di Dieguito?
\\Marzia, passeggiavamo per Lido dei Pini, tu con una fata sul petto ed io con un dragone, quando mi insegnasti a mangiarmi le unghie, ed io ti volevo sposare.
\\Fiammetta, basta il tuo sorriso a rallegrarmi una giornata amara, ci riesci sempre.
\\Andrea, il mio compagno di giochi, da quando mi facevi i dispetti e ti trucidavo nei miei raptus omicidi,
a quando ti unirai al mio vagabondare in America.
\\Trilli, la mia compagna di ballo preferita, sempre giocosa e golosa. Viva la cioccolata.
\\Flami, la mia compagna di pettegolezzo preferita, sveglia e competitiva. Vale cuatro.
\\Nonna, ogni tanto penso che sei l'unica persona con a cui potrei 
confidare le mie turbe bislacche avendo la certezza di essere compreso.
Più passa il tempo e più mi accorgo di essere uguale a te.
Spesso mi soffermo a rimpiangere un confronto da adulti.\\
\begin{center}
Mam\'a.
\end{center}
}
}






\end{document}